\documentclass[12pt,epsf]{article}

\usepackage{amssymb,amsmath}

\newcommand{\bc}{\begin{center}}
\newcommand{\ec}{\end{center}}
\newcommand{\bd}{\begin{displaymath}}
\newcommand{\ed}{\end{displaymath}}
\newcommand{\be}{\begin{equation}}
\newcommand{\ee}{\end{equation}}
\newcommand{\ba}{\begin{array}}
\newcommand{\ea}{\end{array}}
\newcommand{\bea}{\begin{eqnarray}}
\newcommand{\eea}{\end{eqnarray}}
\newcommand{\bt}{\begin{tabular}}
\newcommand{\et}{\end{tabular}}

\newcommand{\bp}{\begin{picture}}
\newcommand{\ep}{\end{picture}}
\newcommand{\bfi}{\begin{figure}}
\newcommand{\efi}{\end{figure}}

\begin{document}

\title{\huge \bf {Tunguska Dark Matter Ball}}

\author{ C. D. Froggatt ${}^{1}$ \footnote{\large\, colin.froggatt@glasgow.ac.uk} \ \ \
H.B.~Nielsen ${}^{2}$ \footnote{\large\, hbech@nbi.dk} \\[5mm]
\itshape{${}^{1}$ Glasgow University, Glasgow, Scotland}\\[0mm]
\itshape{${}^{2}$ The Niels Bohr Institute, Copenhagen, Denmark}}

\maketitle

\begin{abstract}
It is suggested that the Tunguska event in June 1908
was due to a cm-large ball of a condensate
of bound states of 6 top and 6 anti-top quarks containing highly compressed
ordinary matter. Such balls are supposed to make up the dark matter
as we earlier proposed. The expected rate of impact of this kind of
dark matter ball with the
earth seems to crudely match a time scale of 200 years between
the impacts.
The main explosion of the Tunguska event is explained in our
picture as material coming out from deep within the earth, where
it has been heated and compressed by the ball penetrating to a
depth  of several thousand  km. Thus the effect has some
similarity with volcanic  activity as suggested by Kundt.
We discuss the possible identification of kimberlite pipes with
earlier Tunguska-like events.
A discussion of how the dark matter balls may have formed in the
early universe is also given.
\end{abstract}

\date{}


\newpage

\thispagestyle{empty}
\section{Introduction
}

In 1908 a cosmic body is supposed to have fallen down in the Tunguska
region of Siberia \cite{Tunguska}, where its explosion caused the trees to fall
down in an area of 2000 km$^2$. The cosmic body is commonly assumed to have
been a comet or possibly a meteorite. But no genuine piece of a comet or a
meteorite has been definitely identified\footnote{However indirect evidence
from a magnetic and seismic-reflection study \cite{gcubed} of Lake Cheko located
about 8 km NW of the inferred explosion epicentre, is compatible with the
presence of a rocky object 10 m below the lake. It has also been reported
\cite{pravda} that ground penetrating radar studies of the Suslov crater near
the epicentre, by V.~Alexeev and his team from the Troitsk Innovation and
Nuclear Research Institute, indicate the presence of a huge piece of ice buried
deep inside the crater.} in Tunguska.
Many hypotheses about what really happened have been put forward,
but the question is not totally settled.

In the present article we shall propose that what happened in Tunguska was
that a $10^8 $ kg piece of dark matter ran into the earth at
Tunguska. Despite the weight being suggested to be $10^8$ kg in
our proposal, the size of the corresponding dark matter ball is only
expected to have been about a centimetre, $R \approx 0.67$ cm.

It is generally believed that dark matter requires some new physics,
so that it could e.g. consist of the susy-partner of some known neutral
particle. In such models the mass of the dark-matter-particles is of the
order of 100 GeV typically. But really the mass of dark matter particles
is extremely badly determined, because on the large distance scale
the behaviour of the dark matter is independent of how small or how
big are the dark matter particles into which it
is divided.  In fact Frampton et al.~\cite{Frampton} proposed that
the dark-matter particles could be black holes.
In the present article we propose the existence of
dark-matter particles with masses of the order of $10^8$ kg.  Contrary to the
new physics explanations of dark matter, a most important feature
of our model is that we get it formally {\em with the Standard Model alone}.
Not even the forces responsible for non-zero neutrino masses, whatever they
may be, play any significant role in our model. The only new physics
introduced in our model is our requirement that the parameters
of the Standard Model should be {\em fine-tuned}
in a special way, but formally it is just the Standard Model.

We have in fact earlier proposed \cite{crypto} that, by imposing a certain principle of
fine-tuning, there is the possibility for a model of dark matter consisting
of pea-size or hand-size balls of a new type of vacuum in which certain
bound states of 6 top and 6 anti-top quarks are Bose condensed.
Inside their region of the new\footnote{Strictly speaking
this phase is only the new one in the sense that we have known about
the outside phase much longer. We should also note that a model for dark
matter using an alternative phase of QCD has been proposed by Oaknin and
Zhitnitsky \cite{zhitnitsky}.}
vacuum these balls contain some highly
compressed ordinary matter, which is
kept compressed by the skin separating the two phases of the vacuum
(the one with the condensate and the one without).

Our balls are supposed to come out of the early universe
development before the temperature had fallen to about 2.3 MeV
(as we discuss in section \ref{bbn}) and would therefore not
significantly effect standard Big Bang Nucleosynthesis.
Their whole existence is strongly based on the hypothesis that
there exists at least the two types of vacua mentioned above. In
this scenario, at temperatures above the weak scale $\approx 100 $ GeV,
both phases of the vacuum would exist together randomly mixed over space.
However, as the temperature lowers, the walls along the border of the one
phase to the other will contract and presumably at the end only
one phase will survive, unless some mechanism can stabilise small
balls of the surrendering phase. It is of crucial importance that
we assume that in the phase {\em with} the bound state condensate the Higgs
field is somewhat smaller than in the usual phase without the bound state
condensate. This, in turn, makes the masses of the
nucleons somewhat smaller in the phase with the bound state condensate,
so that they get attracted to that
phase. Then it becomes possible for a ball, having the vacuum with the
bound state condensate inside it, to be filled with nucleons or just matter of
the ordinary type and be kept from collapsing. In this way some bit of
the otherwise surrendering phase, which we take to be the one with the
bound state condensate, can survive in balls. It is these balls we want to take
to be the dark matter.
The occurrence of the above two phases of the vacuum, which are so
crucial for our model, is supposed to be due to our postulate
\cite{MPP} of the ``Multiple Point Principle'', according to which
the parameters of any given model should be fine-tuned to ensure
the existence of two or more phases with the same energy density.

In \cite{tophiggs} we applied this multiple point principle (MPP)
to the Standard Model, requiring the effective potential for the
Higgs field to possess two degenerate minima taken to be at the
weak and Planck scales respectively. This vacuum degeneracy
requirement led to the postdiction of $M_t = 173 \pm 4$ GeV for
the top quark mass and the prediction of $M_H =135 \pm 9$ GeV for
the Higgs mass. According to MPP, the values of the top quark and
Higgs masses should lie on the Standard Model vacuum stability
curve \cite{stabilitycurve}. There have recently been
recalculations \cite{Degrassi} of this vacuum stability curve at
next-to-next-to leading order. Using a top quark pole mass of $M_t
= 173$ GeV as input leads to an updated MPP prediction for the
Higgs mass of $M_H = 129$ GeV, with a theoretical uncertainty of
$\pm 1$ GeV. However this result is rather sensitive to the value
of the top quark pole mass: a change of $\Delta M_t = \pm 1$ GeV
gives a change in the MPP predicted Higgs mass of $\Delta M_H =
2\Delta M_t = \pm 2$ GeV. Furthermore there is considerable
ambiguity in the extraction of the top quark pole mass from
experimental measurements using top quark decay products and Monte
Carlo simulations \cite{tevatron-lhc}. Thus, within the
uncertainty \cite{alekhin}, the MPP prediction for the Higgs mass
is in agreement with the 126 GeV mass of the particle recently
discovered at the LHC \cite{lhc}.
We have also considered \cite{metampp} an alternative formulation of MPP,
which we refer to as meta-MPP, in
which the two vacua, with low and high Higgs VEVs respectively, lie on the
borderline of metastability due to thermal vacuum decay in the early universe.
The updated \cite{espinosa}
meta-MPP prediction for the Higgs mass, using $M_t = 173$ GeV as input, is
$M_H = 121$ GeV with a similar uncertainty as for the MPP prediction.

As a further application of MPP,
we have suggested that the parameters of the
Standard Model, in particular the observed top quark Yukawa coupling constant,
are also fine-tuned to give yet
another degenerate phase of the Standard Model vacuum.
This third vacuum has a weak
scale Higgs VEV somewhat smaller than the usual 246 GeV and
contains a Bose condensate of a strongly bound 6 top + 6
anti-top quark state \cite{nbs,boundstate,Kuchiev}. It is only the
two low (weak) scale vacua (with and without the bound state
condensate) that are relevant for the model of dark matter balls
\cite{crypto} which we use in this paper.

We expect there to be a wall or brane-like structure between these
two phases - a solitonic wall - with a tension given by the weak
interaction scale, which is a very high tension compared to the
nuclear physics scale. Especially there must be such a brane-like
structure or skin around the balls that make up the dark matter.
Because of the high tension in the skin the dark matter balls must
have a relatively big size, in order that they shall not totally
quench and allow the ordinary matter inside to spit out into the
outside  vacuum.
Therefore balls below some size limit will contract and have their
with condensate vacuum inside disappear, while balls bigger than
this limit will remain as stable balls.

In our scenario a major part of the nuclear matter, with its
accompanying electrons, is strongly compressed into our balls
before the temperature fell to 2.3 MeV. These nucleons inside the
balls would avoid effectively participating in the standard Big
Bang Nucleosynthesis \cite{Kolb}. Thus the standard fitting of the
ratios of the abundances of the light (helium lithium, deuterium
,...) elements being composed in this era would lead to a ratio of
baryon number to photon number  $\eta \approx 10^{-9}$
corresponding to the number of baryons being {\em outside the
balls} in this era. The point is, of course, that the baryons
inside the balls are effectively being kept aside from the
processes going on in the standard Big Bang Nucleosynthesis time.

It should also  be borne in mind that, even though these balls
will still interact somewhat with ordinary matter at the surface,
their supposed masses of order $10^8$ kg for the rather small
pea-sized balls are so large compared to their size that the
gravitational interaction comes to dominate. In this way they
function as {\em dark} matter.

It is of course important that our dark matter balls indeed have
the right properties for dark matter:

\begin{enumerate}

\item
Our dark matter balls must be stable on cosmological time scales.
Since they actually consist of ordinary matter just strongly
compressed, they are stable simply because of baryon number
conservation. So stability is more natural in our model
than in the usual WIMP models where a special conserved quantum
number, such as R-parity conservation, has to be invented and
added to the theory just to ensure the wanted stability.
As mentioned above the dark matter balls have to be larger than the
critical size of order 0.67 cm, in order
to avoid collapse under the pressure from the skin around the ball.
We should therefore estimate the lifetime of a typical dark matter ball
due to the skin tunneling into the ball and increasing the pressure
so much that the matter inside gets spit out.
Let us consider, for example, a ball with a radius 1 mm larger than
the critical radius. There will be a mass of order $10^7$ kg outside the
critical radius and each nucleon will have an effective binding energy of
order 1 MeV. So the energy violated by the skin tunneling to the critical
radius will be of order the Einstein energy of $10^4$ kg. The corresponding
tunneling time is greater than or equal to the time taken for light to travel
1 mm. It is clear that the tunneling probability for such a process is
completely negligible in the lifetime of the universe.

\item
 The interaction of a dark matter ball with normal matter will, in order
 of magnitude, be similar to that of an ordinary matter ball of the
 same size. But, compared to the mass of the dark matter ball, this is
 a very small interaction. An important feature, revealing the lack
 of interaction between dark matter balls, is the question of whether their
 interaction with say light can cool down a gas of dark matter so that
 it would contract and form stars much like atoms. Now the typical
 distance between the dark matter balls is of order an astronomical unit.
 So the dark matter balls only very rarely come close to one another
 and, consequently, are not accelerated and thus do not radiate.
 Therefore a halo of dark matter balls will not flatten into a disk.

 \item
 Due to their enormous mass the dark matter balls are extremely cold
 in the sense of being non-relativistic, even if having a high temperature.
 The dynamics of dark matter on a large scale is independent of the
 mass of the constituents, provided there are many constituents in the
 region under consideration. This means that, over distances large
 compared to an astronomical unit, our dark matter balls act just like
 the dark matter in the standard $\Lambda CDM$ cosmological model.

 \item
 Although dark matter balls have a mass of order $10^8$ GeV, they are not
 heavy enough to be detected in microlensing searches. However such a
 ball would only hit the earth once every two hundred years or so,
 causing a Tunguska-like event, and will not be seen in the current
 dark matter direct detection experiments. So a positive signal for
 dark matter in these searches would falsify our model. Similarly
 dark matter balls would not combine with known elements to form
 anomalously heavy isotopes.

\end{enumerate}

The main point of the present article is now to discuss how these balls might be
observed via their {\em relatively} smaller non-gravitational interactions.
In fact we suggest that one of these balls hit the earth in Tunguska in 1908
causing the famous Tunguska event \cite{Tunguska}.

In the following section 2 we shall estimate the mass of a dark matter
particle (ball) needed, in order that a ball should hit the earth around once
every hundred years or so, as one must imagine the rate of Tunguska-like events
to be. For this estimation we shall assume that the density of our dark matter
balls reproduces the phenomenological value of the dark matter density.
In section 3 we shall discuss the minimal size of the ball of
the vacuum with the condensate  needed to prevent its collapse and the
liberation of its contained baryons. We observe that this minimal weight or size,
which probably gives a typical dark matter ball mass, matches well with the mass
needed to reproduce the rate of Tunguska events.

In section 4 we estimate, although we can only do so rather crudely,
the attractive potential pulling the nucleons into the phase with
the condensate. Our estimate for the energy gained by a nucleon
going into the with condensate phase comes out to be around 10 MeV.

In section 5 we then explain what in our picture should have happened
in Tunguska.

In section 6 we discuss the possibility that the geological funnels known as
kimberlite pipes, through which kimberlite magmas from deep within the earth
erupt at the earth's surface, were
actually created by earlier Tunguska-like events \cite{Kundt,cudo}.
In section 7 we shall then show that indeed the fall of a ball like
ours would penetrate very deep into the earth, making the event
more reminiscent of a volcanic eruption than the impact of a comet or a
meteor.
In section 8 we discuss how our dark matter balls might really have been
produced in the early universe
before the temperature reached down to 2.3 MeV.

In section 9 we resume and conclude that
our picture of dark matter provides a viable explanation of the Tunguska event.

Finally, in two appendices, we put forward a series of smaller detailed
points about the properties of the bound state condensate and
the cosmological production of our dark matter balls.

\section{Ball mass from Tunguska event time statistics}

Here we assume that the dark matter in the universe consists of balls
of significant weight
and that the Tunguska event was the result of a ``fall down'' of
one such dark matter particle. We can then estimate a mass for
these balls from the presumed time scale between Tunguska events.

From astronomical studies \cite{rhohalo} the density of dark matter in the halo
of our galaxy is
\be
\rho_{halo} = 0.3 \ \hbox{GeV/cm$^{3}$} = 0.3* 10^6 m_N \ \hbox{m$^{-3}$},
\label{halo}
\ee
where $m_N$ is the nucleon mass.

There are reasons though to believe that, at the solar system and in the disk
of the galaxy \cite{Read}, the density of dark matter will be a bit increased
compared to the halo density by something like a factor of 2. So, in the
following, we shall take the dark matter density in the solar system
to be $\rho_{\odot}=2\rho_{halo}$.

We assume that the halo dark matter particles have a typical speed of 220 km/s and that
the dark matter particles which are attached rather to the disk or the solar
system have a typical speed of 60 km/s. So we take a typical speed of
\be
v = \sqrt{\frac{1}{2} 220^2 + \frac{1}{2}60^2}\  \hbox{km/s} =
1.6 *10^5\ \hbox{m/s} \label{speed}
\ee
for the balls making up the dark matter. The two contributions are taken crudely to
be equal, in correspondence with our estimate of the density near the earth being
just double that of the halo proper.
Now the cross-section of the earth is
\be
A_{\oplus} = \pi R_{earth}^2 = \pi * (6.4 * 10^6)^2\ m^2=1.29 * 10^{14}\ \hbox{m$^2$},
  \ee
and so the volume of space that is tested per time unit for the presence of dark matter
by collisions with the earth is thus
\be
V_{check} = v * A_{\oplus} =
2.1 * 10^{19}\ \hbox{m$^3$/s}.\label{Vcheck}
\ee
Therefore on average the mass of dark matter hitting the earth per unit time is
\begin{eqnarray}
V_{check}*2\rho_{halo}
& = & 2.2*10^{-2}\  \hbox{kg/s} =6.9 *10^5\ \hbox{kg/y}.
\end{eqnarray}
Now the time which has elapsed since the Tunguska-event is close to 100 years.
If we want to be a bit more precise and say that the fact that the Tunguska
event occurred 100 years ago rather means that we are approximately in the middle
between two successive events,
then we should take the rate of fall of dark matter on the earth to be
\be
r_B = \frac{1}{2 *100 \ \hbox{y}} = 1.5 *10^{-8}\ \hbox{s$^{-1}$}.\label{twoh}
\ee
The mass of the dark matter balls falling on the earth must thus be estimated to be
\begin{eqnarray}
m_B& =& r_B^{-1} * V_{check}*2\rho_{halo} \\
& =& 1.4*10^{8}\ \hbox{kg}
=1.4*10^5\ \hbox{Tons}. \label{mB}
\end{eqnarray}
This corresponds to the ball containing
\be
m_B/m_N = 1.4 * 10^8 / 1.67 * 10^{-27} = 0.84*10^{35}  \ \hbox{nucleons}.
\label{mbomn}
\ee
The total kinetic energy of the ball on entrance into the earth's atmosphere
will thus be
\be
T_v = \frac{1}{2} m_B *v^2 =
\frac{1}{2}1.4*10^{8}\ \hbox{kg} * (1.6*10^5\ \hbox{m/s})^2 =1.8*10^{18}\ \hbox{J},
\label{Tv}
\ee
which is equivalent to 430 megaton of TNT.

The estimated energy output observed in the Tunguska event is
of the order of 10 to 30 megaton of TNT, which is smaller by a factor of
20 than our above estimate (\ref{Tv}).
As we shall see in more detail, dark matter in our model
consists of cm-large balls of very high mass for their size, say of
the order of our estimate $m_B = 1.4 * 10^8$ kg in equation (\ref{mB}).
This means that the particle in our model penetrates
through the atmosphere and deep inside the earth.
Thus a major part of the kinetic energy of the in-falling dark matter
ball will be dissipated underground, while only a minor fraction
may come out of the earth similarly to volcanic activity.
That the fraction of energy coming out of the earth should be about
1/20 of the total impact energy seems not at all unreasonable.

\subsection{Space between dark matter balls}

We here estimate the volume of space which on average contains just one dark
matter ball, using our above estimate for its typical mass. In the solar system,
where the dark matter density is
$\rho_{\odot} = 2\rho_{halo} = 0.6 \ \hbox{GeV/cm$^{3}$}$,
the volume of space containing just one dark matter ball on average is given by
(\ref{Vcheck}) and (\ref{twoh}) to be
\begin{equation}
 \hbox{``Volume per ball''}|_{\odot} = V_{check} * r_B^{-1} =
1.3*10^{29}\ \hbox{m$^3$}.
\end{equation}
However the present average dark matter density over the whole
universe is \cite{pdg}
\begin{equation}
 \rho_{universe} = 1.2*10^{-6}\ \hbox{GeV/cm$^3$}.
\end{equation}
Therefore the average volume containing one dark matter ball in
the universe today is
\begin{equation}
 \hbox{``Volume per ball''}|_{universe} =
 V_{check} * r_B^{-1} * \rho_{\odot}/\rho_{universe}
 = 6.8*10^{34}\ \hbox{m$^3$}.
\end{equation}
The radius of a sphere with this volume has the length
\begin{equation}
 l = 2.5*10^{11}\  \hbox{m},
\end{equation}
which gives the order of magnitude of the typical distance between
the balls. This distance corresponds to the present temperature of
the universe $T_0 =2.725\ \hbox{K} = 2.35*10^{-4}\ \hbox{eV}$. To
the approximation that the linear size (=scale factor) of the
universe would vary like the inverse temperature $T^{-1}$ the
corresponding distance at temperature $T$ will be
\begin{equation}
 l_T = 2.5*10^{11}\ \hbox{m} * (T_0/T) =
 6\ \hbox{m}\ *\left(\frac{10\ \hbox{MeV}}{T}\right).
\label{lT}
\end{equation}
So, for example, at the 10 MeV temperature era the typical
distance between the balls would be of the order of 6 m.

\section{Our ball-parameters}
It is the basic idea of the present work that the dark matter is made up
of small pieces of another type of vacuum with a $6t + 6\overline{t}$
bound state condensate into which the nuclei are pulled,
because the Higgs field is lower in this other vacuum inside these pieces.
Following the ``Multiple Point Principle'' \cite{MPP}, we shall assume that
the vacua inside and outside the balls have the same energy density.
There is, however, still an appreciable energy density on the
surface between the two different phases. We denote the tension
or energy density per unit area in this surface by $S$. If we now have a
spherical ball of radius $R$ surrounded by such a  surface, it will
provide a pressure $P$ on the interior of the ball (relative to the
pressure in the outside, which in the absence of any matter
is negligibly different from the inside
pressure, due to the assumed equality of the energy densities)
\be
P = 2S/R.\label{s2oR}
\ee

\subsection{Introduction to pressure estimation}
\label{pressure}

In order to prevent the surface of the ball from contracting and
quenching the ball, it is necessary for the contents of the ball
to provide a pressure which supports the ball. This pressure is
dominantly provided by the degeneracy pressure of the essentially
normal matter filling the interior and under enormous pressure
from the surface.

The crucial parameter for estimating the pressure obtainable is the
change in energy per nucleon by passage through the surface of the ball
\be
\hbox{``potential shift''} =\Delta V.
\ee
In section \ref{qmd} we shall discuss the estimation of $\Delta V$,
but here we take as a reasonable value
\be
\Delta V = 10 \pm 7 \ \hbox{MeV}.
\ee

We imagine that the balls, which are supposed to be formed
in the early universe, say around
1/10 of a second after the start, are just barely stable against
collapsing and thereby spitting out their nucleons (and their
electrons). At least the balls must be able to support themselves from
collapsing and so the stability border against collapse must function as
a lower limit for the size of the balls.

In section \ref{early-universe} and appendix B we
shall discuss our scenario for the formation of the balls together with
their nucleon content in the early universe. However let us here just
suppose that the balls started out at the weak scale temperature
$T \approx 100$ GeV as very extended objects, containing the
vacuum with a bound state condensate.
We argue that they will tend to carry the neighbouring plasma along with them.
Thus, in first approximation, one would expect them to follow the Hubble
expansion. But, of course, the tension in the wall tends to make them
diminish in size relative to this Hubble expansion. However, in appendix
\ref{survival}, we put forward an effect which makes sufficiently big balls
expand even more than the Hubble expansion.
Furthermore we estimate, in appendix \ref{fadingout},
that the inertia of the balls is very important and that the balls can
easily collide with each other, thereby causing a rather complicated
motion (at least for temperatures less than 2 GeV).

At the end, the stability of the ball depends on
whether or not it collects sufficiently many nucleons as it contracts to
about the present day size. If there are too few nucleons inside,
then at some point the radius of the ball becomes small enough that
the pressure from the skin of the ball forces the nucleons out and the
ball collapses completely. The number of nucleons inside the ball
of course mainly depends on the original size of the ball, although it
could collect some of the nucleons spit out by collapsing smaller balls.
We expect the balls containing more nucleons than those of
the critical size on the borderline of stability to survive\footnote{However
we also speculate that explosive nuclear fusion in the highly compressed
matter may lead to the expulsion of a fraction of the nucleons.}
as balls. In any case there must be some reasonably smooth distribution
of ball sizes. There is a cut-off at small size given by borderline stability
and at large size \cite{Kibble} given by the Hubble radius $1/H$ at the
time when the nucleons start to be collected inside corresponding to a
temperature of $T \sim 10$ MeV. Suppose we make a power law ansatz for the
smooth mass distribution for the number density of balls $N_{balls}(M)$.
Then, in order for the expression for the total amount of mass per unit
volume to converge at the small and large size cut-offs, we require that
\be
dN_{balls}(M) \propto \frac{\mathrm{d}M}{M^2}. \label{nball}
\ee
This distribution gives
\be
MdN_{balls} \propto \frac{\mathrm{d}M}{M} \label{mnball}
\ee
and the expression for the total amount of mass diverges only
logarithmically at the upper and lower limits for the ball size.
It follows from (\ref{mnball}) that the typical mass for a ball
will be close to the lower limit, which is just the mass on the
stability borderline.

\subsection{The degenerate electrons}
For the stability in time of the ball, the pressure exerted by the material
inside the ball must balance the pressure (\ref{s2oR}) from the wall tension.
Of course this pressure $P$ must be in conformity with the material
properties of the stuff inside the ball.
In fact the dominant contribution to the pressure arises from
highly degenerate electrons in matter much like white dwarf
material. For white dwarf-like material the electrons form a
Fermi-sea up to a Fermi momentum $p_f$. Supposing that this matter
is cold compared to the Fermi energy level - and that is certainly
the case for our balls - the energy density of the electrons is
given as
\be
n_e <E_e>= 2\int_0^{p_f} \sqrt{p^2 + m_e^2} 4\pi p^2 \,
\mathrm{d}p/(2\pi)^3 \approx \frac{1}{4 \pi^2} p_f^4,
\ee
and the density
of electrons is
\be
n_e =2\int_0^{p_f} 4\pi p^2 \, \mathrm{d}p/(2\pi)^3
\approx \frac{1}{3 \pi^2} p_f^3. \label{ne}
\ee
Here $<E_e> $
denotes the average energy of an electron in the Fermi gas and is
\be
<E_e> = 3p_f /4.
\ee
Now the pressure $P$ of a highly
relativistic degenerate electron gas is given by one third of its
energy density and hence
\be
P = <E_e> n_e/3 \approx \frac{1}{12 \pi^2} p_f^4 \label{P}
\ee
Assuming that the nucleons occur roughly as an equal amount of
neutrons and protons, there must be about twice as many nucleons
as electrons,
\be
n = 2 n_e.\label{nne}
\ee

\subsection{The Critical Ball Parameters}

We now want to consider the ball size just to be on the borderline
of stability against collapse because, as argued in subsection
\ref{pressure}, we expect that the typical ball size produced in
the early universe to be close to this border for collapsing.
Consider indeed a ball just about to spit out its nucleons
together with associated electrons. We may consider the energy
required to release an electron and its associated proton-neutron
pair as an effective potential term for the electron. For a ball
on the stability borderline we require that such a formal
electron, which is really an electron associated with its two
nucleons, just has sufficient energy to escape and become ``free''
in the vacuum outside the ball. If we take the potential
associated with the nucleons to be zero outside and -$2\Delta V$
inside the ball, the energy of the combined electron and
proton-neutron pair outside is $m_e$. In the critical ball case,
we must therefore have that the energy of the combination inside
$-2\Delta V + E_e $ must be equal to $m_e$ for those electrons
that can just go in and out of the ball. This means that the
Fermi-surface energy minus the rest energy for the electrons must
obey
\be
E_{ef}-m_e = 2\Delta V.
\ee
Or crudely ignoring the
rather small electron mass $m_e$ compared to $2\Delta V$ and
noting that the electrons turn out here to be highly relativistic,
we simply have
\be
p_f = 2\Delta V. \label{pf}
\ee

Substituting the above value (\ref{pf}) for the Fermi-surface
momentum into (\ref{ne}), we obtain
\be
n_e = (2\Delta V)^3 / (3 \pi^2),\label{e31}
\ee
which, from equation (\ref{P}), gives
\be
P = (2\Delta V)^4 / (12 \pi^2) = \frac{4}{3 \pi^2} (\Delta V)^4.
\label{pn2}
\ee

Combining (\ref{pn2}) with (\ref{s2oR}) we get
\be
2S/R = \frac{4}{3\pi^2} (\Delta V)^4.
\ee

\subsection{Connecting to the mass}

The mass of an on borderline stable ball of ours is given as
\begin{align}
m_B|_{border}& = \frac{4\pi }{3} * R^3 * m_N * n \\
 & = \frac{4\pi}{3} \left (\frac{2*3\pi^2S}{4 (\Delta V)^4}\right )^3 *m_N
\frac{2*(2\Delta V)^3}{3\pi^2} \\
& = \left ( \frac{S}{(\Delta V)^3} \right )^3 * m_N * \pi^5 *3*2^{3} \\
& \approx 7.3 * 10^3 \left (\frac{S}{(\Delta V)^3} \right )^3 * m_N.
\end{align}

From here we deduce
\be
 m_B/m_N = 7.3 * 10^3 \left (\frac{S}{(\Delta V)^3} \right )^3. \label{mbdv9}
\ee
Then, using equation (\ref{mbomn}) based on the expected rate of Tunguska
events, we have
\be
\left (\frac{S}{(\Delta V)^3} \right )^3 = \frac{0.84 * 10^{35}}{7.3 * 10^3}
= 1.14 * 10^{31}
\ee
leading to
\be
\frac{S^{1/3}}{\Delta V} = (1.14 *10^{31})^{1/9} = 2.8 *10^3.
\ee
Thus, with $\Delta V = 10$ MeV,  we get the cubic root of the tension $S$ to be
\be
S^{1/3} = 28 \ \hbox{GeV}. \label{e42}
\ee

In appendix \ref{vacuumstructure} we shall estimate the expected
value of this tension in the border between the two vacua, as well
as we can, from the properties of the bound state condensate.
This estimate turns out to be $\sim 16$ GeV, which is
in reasonable agreement with ({\ref{e42}). The crux of the matter
is that this quantity $S^{1/3}$ is supposed to be related to the
top quark mass, which determines the radius of the bound state
\cite{boundstate}, or be somehow given by the weak
interaction scale, which would typically be counted as 100 GeV.
In any case the order of magnitude $S^{1/3} \sim 28$ GeV, gotten from the rate of
Tunguska events and the supposed Higgs field suppression in the vacuum
inside the ball, agrees well with the estimates in appendix \ref{vacuumstructure}.

Because of the ninth root under which the rate of Tunguska events
occur in our calculation, a factor of two uncertainty in the cubic
root of $S$ would correspond to a factor of 512 in this rate. It
would certainly be unreasonable to think that we got a Tunguska
event a hundred years ago, if such events only occurred every
$51200$ years say. So the rate of Tunguska events is more
accurately estimated for our purpose than the quark mass change
effect $\Delta V$ which is uncertain, according to our estimate in
section \ref{qmd}, by about a factor of 2. Also of course the
``weak scale'' only makes sense to be used in dimensional
arguments up to a factor of $\sim 2$ uncertainty.

\subsection{Combined information on ball size}
\label{sec:ball-size}
The agreement of our fitted value (\ref{e42}) of the tension
with the weak scale
provides phenomenological support for our picture and the hypothesis of
the balls being on the border of collapsing. So this gives us confidence in using
formulas like (\ref{nne}, \ref{e31}) with $\Delta V = 10$ MeV
to calculate the ball mass density:
\be
\rho_B = m_N * n = 2m_N * \frac{(2\Delta V)^3}{3 \pi^2} = 5.08 * 10^5\ \hbox{MeV}^4
= 1.13 * 10^{14}\  \hbox{kg/m}^3.
\ee
With our mass estimate  (\ref{mB}) this gives the volume of the ball to be
\be
V_B = \frac{m_B}{\rho_B} = \frac{1.4 * 10^8 \ \hbox{kg}}{1.13 *10^{14} \ \hbox{kg/m}^3}
= 1.24\ \hbox{cm}^3.
\ee
This ball volume corresponds to a ball radius of
\be
R = (3V_B/(4\pi))^{1/3} =
0.67 \ \hbox{cm}. \label{crudeR}
\ee

\section{Estimation of attractive potential of nucleons to the
vacuum inside the balls}
\label{qmd}
The philosophy of our model is that the region inside the balls has a different
vacuum containing a $6t + 6\overline{t}$ bound state condensate. In this
vacuum with a bound state condensate, the Higgs field is expected
to be smaller than in the vacuum outside the balls by some factor of order unity, say
half as large as in the usual vacuum without a bound state condensate.
This means that the masses of hadrons,
such as the nucleons, will be changed in going from one vacuum to the other one.
Since the Higgs field expectation value is reduced inside the $6t + 6\overline{t}$
bound states \cite{boundstate}, we naturally must have the lower average Higgs field
in the vacuum with a bound state condensate relative to the one without a bound
state condensate. We refer the reader to appendix \ref{vev2} for more details.
The masses of the nucleons are crudely expected
to have additive contributions given by the contained (valence) quarks.
So we expect the mass contribution to the neutron and proton from the quarks
to be
\begin{align}
m_n|_{valence \ quarks} & = 2m_d + m_u = 17 \ \hbox{MeV}\\
m_p|_{valence \ quarks} & =  m_d + 2 m_u = 14 \ \hbox{MeV}
\end{align}
respectively. Here we used quark masses \cite{pdg} renormalised at 1 GeV
\be
m_d = 1.35 * 5.05 \ \hbox{MeV} = 6.82 \ \hbox{MeV}; \ \ \ m_u = 1.35 * 2.49 \ \hbox{MeV}
= 3.36 \  \hbox{MeV}.
\ee
The average of these two contributions, i.e.~the average valence quark
contribution to the nucleon mass is about
\be
m_N|_{valence \ quark}   = 15 \ \hbox{MeV}.
\ee
This means that if, for example, we have a half as large Higgs field expectation
value in the vacuum with a bound state condensate as in the vacuum without a
bound state condensate (the one we live in),
the nucleon energy inside the vacuum with a bound state condensate will be lowered by
15/2 MeV = 7.5 MeV. That is to say there will be an effective
negative potential for nucleons in the vacuum with a bound state of size 7.5 MeV.

Now, however, we expect that the masses of the pions, which are
Nambu-Goldstone bosons, are very sensitive to the quark masses which,
in turn, depend on the Higgs field VEV and vary appreciably from
phase to phase (vacuum to vacuum). In the vacuum with a bound state
condensate, the quark masses are  smaller and
so also the pion masses are smaller there. This in turn makes
the binding of the nuclei stronger and thus the nuclei should
be more strongly kept inside the vacuum with a bound state condensate
than the nucleons when unbound. Very crudely we take it that the main change
in the binding energy of nucleons into nuclei, due to the change
in the Higgs field VEV and thereby the quark and pion masses, is
given by the shift in the range of the pion exchange force
(Yukawa potential). We obtain the relative change in this Yukawa
potential by considering the Yukawa exponential factor
$\exp(- m_{\pi} r )$, where $r$ is the distance between the nucleons
binding to each other and $m_{\pi}$  is the pion mass.
The typical distance between neighbouring nucleons in a nucleus
may be estimated from the semi-empirical picture of the nucleus, in which
the radius $R$ of the nucleus is related to the number density by
the formula
\be
n = \frac{3A}{4\pi R^3} = 1.2 *10^{44}\ \hbox{m}^{-3}.
\ee
Here $A$ is the number of nucleons in the nucleus and, thus,
we get for $A=1$ a radius for a single nucleon formally
\be
R|_{A=1} = (1.2 *10^{44}\ m^{-3} *4 \pi/3)^{-1/3} = 1.26 *10^{-15} \ \hbox{m}.
\ee
This means that the typical distance between neighbouring nucleons in a
nucleus is about twice this quantity, i.e.~$ \hbox{``distance''}
\approx 2 * 1.26\  \hbox{fm} = 2.5 \ \hbox{fm}$. The pion Compton wavelength
is $m_{\pi}^{-1} = (140 \ \hbox{MeV})^{-1} = 1.4 \ \hbox{fm}$.
We now assume that the quark masses inside the vacuum {\em with} a bound
state condensate are lowered by a factor of 2, so that the pion mass $m_{\pi}'$ in
the phase with a condensate is reduced by a factor $m_{\pi}/m_{\pi}' =\sqrt{2}$.
It follows that the pion Compton wavelength increases from 1.4 fm to 2 fm.
That is to say that
the ratio of the pion mass dependent exponential factors, in the phase with
a bound state condensate
relative to the usual one without a bound state condensate, at this typical distance is
\be
\frac{\exp(-\hbox{``distance''} * m_{\pi}')}{\exp(-\hbox{``distance''} * m_{\pi})} =
\frac{\exp(-2.5/2.0)}{\exp(-2.5/1.4)} =
1.7.
\label{pionexchange}
\ee

Assuming that the shape of the potential is not changing too much
and that the kinetic energy of the nucleons inside the nuclei is given by
some sort of virial theorem (like the kinetic energy of the electron in
the atom is given by a virial theorem to be just half the potential
energy), we would expect the kinetic energy to be proportional to the
potential energy.
If the potential were pion exchange dominated, this would mean that the
kinetic energy of the nucleons inside the nuclei should be increased by
the same factor (\ref{pionexchange}) of 1.7. Let us assume, however, that
approximately half of the binding energy is due to pion exchange.
Then we obtain a correction to just half the binding energy by a factor of
1.7 in going into the phase {\em with} a condensate. Now the nuclear binding energy
per nucleon in the usual phase without a condensate is about 8 MeV.
So half of this binding energy per nucleon, i.e.~4 MeV, is changed to $1.7 * 4$ MeV
in the phase with a condensate. In other words the binding energy per nucleon is
larger by an amount $0.7 * 4 \approx 3$ MeV in the phase with a condensate.

Due to quantum fluctuations, the effective typical distance for
the pion exchange interaction could well be smaller, say by $25
\%$, than 2.5 fm. This would lead to a reduction in the correction
factor (\ref{pionexchange}) from 1.7 to 1.4, giving an extra
binding energy per nucleon in the phase with a condensate of only
2 MeV.

Our estimate of 2 or 3 MeV for the extra binding energy per nucleon in the
phase with a bound state condensate makes the effect of the nuclear
binding energy smaller than the effect of the quark masses on the
nucleon masses.  So the large uncertainty in our estimate of the
change in the nuclear binding effect is not so important
for the estimate of the total attraction of ``bound nucleons ''
into the vacuum with a bound state condensate.

In order to have a nice standard value for the attractive
potential driving the nucleons, when bound inside nuclei, into the
vacuum with a bound state condensate, we shall take $\Delta V =
10$ MeV $\approx 7.5 + (2 \ \hbox{or} \ 3)$ MeV as a reasonable
estimate. However we really only have a guess for the order of
magnitude of the reduction of the Higgs field in the vacuum with a
bound state condensate\footnote{In appendix \ref{vev2} we make a
crude estimate of this reduction in the Higgs field using the
properties of the bound state condensate and obtain the ratio of
the Higgs field VEVs in the two vacua to be of order 0.3. This
result is consistent with the value of 0.5 used here and
throughout the paper.}. Thus we clearly cannot avoid an
uncertainty, even for the changes in the nucleon masses, of the
order of a factor of 2 or so. So let us, in conclusion, say that the
binding energy of a nucleon is increased in the vacuum with a
bound state condensate by somewhere between $ 4 \ \hbox{MeV} $ and
$ 18 \ \hbox{MeV}$, which we can write as $\Delta V = 10 \pm 7 \
\hbox{MeV}$. A better estimation of this quantity would of course
be highly desirable, since the mass of our dark matter ball
(\ref{mbdv9}) is inversely proportional to the ninth power of
$\Delta V$. Consequently an uncertainty by a factor of two in
$\Delta V$ corresponds to an uncertainty by a factor of $2^9 \sim
500$ in the mass.

The above analysis of the change in the nuclear binding energy per
nucleon due to a reduction in the quark mass by a factor of 2 is
of course rather simplistic. A more detailed numerical analysis of
the dependence of the binding energy of light nuclei on the quark
mass has been made by Flambaum and Wiringa \cite{Flambaum}.
Extrapolation of their results to the case of a reduction in the
quark mass by a factor of 2 leads to an increase in the binding
energy per nucleon by a few MeV. However, although agreeing in
order of magnitude with our crude estimate of the change in
binding energy, one pion exchange is not the dominant term in
their calculation. Rather they find that two pion exchange
provides an appreciably bigger change in the binding due to the
quark mass variation than does single pion exchange. In addition
vector meson exchange and other contributions provide a change in
the binding of the opposite sign and somewhat compensate the two
pion exchange contribution. Combining all the contributions,
Flambaum and Wiringa obtain values for the fractional change in
binding energy E relative to the fractional change in quark mass
$K = \frac{\delta E / E}{\delta m_q / m_q}$ for nuclei having $A =
3 \hbox{ to } 8$ in the range $K= -1 \hbox{ to } -1.5$. These
light nuclei have binding energies of order 7 MeV per nucleon and
we choose to reduce the quark mass by a factor of two $\delta m_q
/m_q = -1/2$. Thus from reference \cite{Flambaum}, we obtain an
extra nuclear binding energy per nucleon in the vacuum with a
bound state condensate of $\delta E \approx (-1 \hbox{ to } -1.5 )
*(- 0.5) * 7 \hbox{ MeV}
 \approx 4 \hbox{ MeV}$.
This result is accidentally not so far from our crude estimate of
2 or 3 MeV. In any case it is smaller than the change in the
nucleon mass of the order of 7.5 MeV in the vacuum with a bound
state condensate and does not significantly effect our estimate
$\Delta V = 10 \pm 7$ MeV of the potential shift between the two
vacua.

\section{What happened in Tunguska?}
The main idea of the present article is that what happened in Tunguska
in July 1908 was that one of our dark matter balls hit the earth with a
speed (\ref{speed}) of order 160 km/s.
With its large mass $m_B = 1.4 * 10^8$ kg (see (\ref{mB})) and very small size
- having a radius of about $R = 0.67 \ \hbox{cm}$ (\ref{crudeR}) - the dark
matter ball could not at all have been stopped near the surface of the earth
like a meteorite, let alone in the air as it is speculated for a comet.
Rather it would penetrate deeply into the earth and
only get stopped after say a few thousand kilometres.

The kinetic energy of such a dark matter ball would be
$T_v = \frac{1}{2} m_B  v^2 \approx 1.8 *10^{18}\ \hbox{J}$.
It would deliver some small amount of this kinetic energy in the air
presumably making its track visible, if somebody had looked; but by far the major
part of this kinetic energy would be delivered deep inside the earth. Therefore
the main effects of the fall of our ball should be caused by energy coming
out of the earth rather than as if coming directly from an extraterrestrial object.
Hence, to first approximation, our hypothesis would
simulate a {\em volcanic explosion}, which Kundt \cite{Kundt} has suggested
to be the cause of the Tunguska event
rather than the fall of a meteorite or comet.
To first approximation the ball will simply push and heat up enormously
a tube through the earth not very large in radius at first compared to
the cm-size of the ball. One must then imagine that at least part of the
heat energy from this thin tube will push back heated earth material
along the immediately formed tunnel. This tunnel would presumably lie
in a skew direction, since it is very unlikely that the ball should have
hit just head on into the earth. Thus  a lot of heated ``volcanic matter''
would be sent out of the earth in a skew way. It is, however, quite
likely that some of the heated material further down the originally formed
tube would find a shorter way out and up to the surface of the earth than
just back along the tube. The heated material could indeed go more
vertically up towards the surface of the earth just above the place where
it was formed from the passage of the ball. This could mean that several
tunnels would be formed from the track of the ball deep inside the earth
up to the surface, along which the earth material would be pushed up into
the air above the surface of the earth. Indeed Kundt \cite{Kundt} has suggested
that the Tunguska event corresponded to the formation of kimberlite pipes,
which are tunnels of this type that become carrot shaped near the
surface \cite{kimberlite}.
They are usually supposed to come from  ``volcanic explosions'' hundreds
of kilometres under the surface. However we suppose that pipes of this type
were created at Tunguska by the hot tube formed by the passage of the ball.
In section 6 we shall consider the possibility that all other kimberlite
pipes were created by earlier Tunguska-like events.

The presumably gaseous material coming out of these tunnels would go
up in the air, spread out and give rise to the observed explosion in the air
above Tunguska. This would create winds and shock waves, causing trees
to fall and the branches to be ripped off some of the trees left standing
to form the so-called telegraph poles.
We expect only a fraction of the energy of the incoming ball to be available
to participate in the observed explosion, since certainly a part of the energy must
remain inside the earth. Especially if the ball penetrated so deep that its
heated material cannot find a way out to the surface, an appreciable part of
the energy would have to remain inside the earth.

We imagine, as our most likely picture, that the impact of our ball took place in Lake Cheko,
and that some of the funnels observed by Kulik \cite{Kulik} are outburst places
through which hot material found a shorter way out to the surface than simply
going back along the first formed tube. The Suslov Crater, which is near the epicentre
of the Tunguska explosion, we suppose is
one of these places where outburst occurred from the earth via a shorter route.

\subsection{How did the ball fall?}
There are strong indications \cite{gcubed,cheko} that Lake Cheko is the site where
the extra-terrestrial object hit the solid surface of the earth, and we shall
assume that to be the case. The epicentre of the explosion, as indicated
by the tree falling pattern, is however 10 km to the south-east of
Lake Cheko. So, if the Tunguska event were caused by a comet or a meteorite,
the impact should have come from the south-east and have been directed towards
north-west. We however, must have a different picture if we want to keep
to Lake Cheko as the point of impact. Since our ball would cause the main explosion
after already penetrating several kilometres inside the earth, the epicenter
must lie later along the route of the ball in our model. That is to say that
we need the ball to have come from the north-west and oriented towards the
south-east (opposite to the direction needed for the comet or meteorite).

In this picture then the first spit out of hot material back
through the tunnel, formed as the ball passed through, would go out
into the air from Lake Cheko in the opposite direction to the
motion of the ball itself. That is to say the spit out material
would come out in the direction towards the north-west. It is quite
likely that this material would have had very high velocity and
that some of it might be close to or have gone into orbit around
the earth. Thus it is predicted in our model that material from
the interior of the earth was pushed out with a velocity like a
satellite moving in direction towards the north-west. Such material
would quickly reach Britain and Denmark without any mysterious
need for two-dimensional turbulent flow in the upper atmosphere.
Consequently, such a blow-out in the  north-west direction has a
chance to explain the very rapidly occurring noctilucent night
clouds in Northern Europe after the Tunguska event.

We imagine that the major part of the kinetic energy of the ball was dissipated
deep in the earth, say of the order of thousands of kilometres deep. The heated
material from deep within the earth then began to move backwards along the
tunnel made by the passage of the ball. This material would move slower than
the original speed of the ball and presumably make more noise. A major blob of
material moving in this way backwards along the tunnel would cause a loud sound
that would have been conceived, by the witnesses to the Tunguska event, as an
object moving in the opposite direction to the original motion of our dark matter
ball. So the witnesses would hear the sound moving from the south-east to the
north-west. It is claimed that they described it as moving from east to north
\cite{wiki}.

\subsection{Most important puzzles}
The most remarkable fact about the presumed in-fall in Tunguska is
that, in spite of the large energy dissipation in the atmosphere
equivalent to $ 10 \ \hbox{to} \ 30\ \hbox{MTons}$ of TNT, there
were no crater and no left over material from the large in-falling
object. In our model this remarkable lack of material from the
in-falling body is explained by the body being so small compared
to its mass that it could not be stopped and dug itself so deep
into the earth - and perhaps even slung out from the other side of
the earth - that nobody could find it yet.
Rather than a crater we predict the existence of funnels created by
material expelled from the earth itself
and indeed that is closer to what Lake Cheko seems to be, an opening of a
kimberlite-like funnel.

The trees fell oriented as from a centre and not as if the
velocity of the in-falling object had given the wind a main
direction. In our picture the explosion came from the interior of
the earth similar to what happens in Kundt's \cite{Kundt}
volcano-like model for the Tunguska event. Actually we have to
imagine that the  seeming explosion in the air came from the earth
by material, gaseous or solid whatever, coming out through the
funnels that Kulik \cite{Kulik} saw in the region. At least a
picture in which the explosion is replaced by a volcano-like
emission from the earth, rather than a fast moving object creating
the air motion directly, has a much better chance of delivering a
pattern of trees falling to different sides than a fast passing
object with its strong directional effect.

Also the ``telegraph poles'' - trees stripped of their bark and
branches but still standing - would more easily appear in an
explosion having its origin as coming out of the earth than in one
produced directly by something having a high speed along the
earth.

The problem with the small amounts of iridium measured
\cite{Rasmussen, Kolesnikov1} in the layers of the peat formed in
1908 in Tunguska is a major reason for not believing that it was a
meteorite which fell down in Tunguska. But even for the hypothesis
that it was a comet, the iridium content seems low; at least it
could not be a comet with the same dust to ice composition as the
Halley's comet, which has approximately 40\% dust. Really to
explain the carbon to iridium ratios a comet with an almost pure
ice core with admixtures of hydrocarbons and other organic
compounds is needed \cite{Kolesnikov2}. In our model the extra
material from the catastrophe year comes from the earth
underneath, perhaps from the mantle of the earth since our ball
likely went extremely deep. But presumably the iridium of the
earth has sunk even deeper than the mantle and so, contrary to
dust in comets or meteorites which are rich in iridium, the mantle
or material closer to the surface of the earth will have
comparably little iridium content. In this way our model, with
stuff expelled from the earth, would easily explain the major
mystery of why so little iridium has been found in the Tunguska
peat.

A depletion of radioactive carbon 14 and an excess of carbon 13 are also
observed \cite{Rasmussen, Kolesnikov1} in the Tunguska peat
samples of 1908. In our model
the event is mainly dominated by the eruption of materials
already in the earth and not by
extra-terrestrial materials. Some outburst from the deeper layers
of the earth would be depleted in carbon 14 which is formed in the
atmosphere. This could well lead to the observed drop of one or two percent in
the $^{14}C$ content of the peat. The standard measure of the ratio of
the stable isotopes $^{13}C$ and $^{12}C$ is $\delta^{13}C$, which
is defined as the parts per mille difference between the $^{13}C$
content of the sample and that of the so-called PDB standard.
The main organic component in the peat has $\delta^{13}C \approx -26$,
while the peat found in the layers around the Tunguska event shows a
shift \cite{Rasmussen, Kolesnikov1} of about +2 parts per mille
in $\delta^{13}C$. Now carbon with $\delta^{13}C \approx -5$ is a
major component of the earth's mantle \cite{Deines}. So the
admixture of mantle material in the peat could lead to an
increase of 2 parts per mille in $\delta^{13}C$.

\section{Kimberlites}
In this section we investigate the hypothesis that kimberlite pipes
\cite{kimberlite} are created by Tunguska-like events.

Kimberlite pipes are
funnels of rocks seemingly coming from very deep in the
earth and sometimes carrying diamonds. These pipes
commonly occur in clusters of less than 50 pipes of variable extent.
The pipes rarely occur at distances larger than 50 km from the
cluster field. The age ranges from Archean to Tertiary times, and kimberlite
emplacements of different ages can occur in the same location.
Based on presently available information, the kimberlites are restricted
to the Continental Intraplate settings.

The idea, which can be said to be due to Kundt \cite{Kundt}, is that
a Tunguska type event is connected with the formation of
kimberlite pipes and material being expelled through them.
So that the Tunguska event is indeed
a volcanic event. We ourselves propose that a cluster of
kimberlite pipes is formed as a result of the impact of
a dark matter ball. This ball penetrates
into the earth and creates extremely strongly heated
material deep in the earth, which then finds some ways out to give the observed explosion
by forming the various pipes in the cluster. It is comfortable for our theory that
the order of magnitude of the distance from the cluster of pipes where no
more pipes are found - the 50 km - is much smaller than the ball penetration depth
estimated in section \ref{penetration}. This estimated depth
$\approx $ 1700 km is likely to be uncertain by about an order of magnitude;
so the ball would be deposited deep inside the earth or possibly even pass
straight through.

The restriction to the intraplate regions is a priori not so welcome
for our theory, in as far as we expect to find equally
many falls of dark matter balls on any unit area of the earth surface.
The position of the fall must be determined by quite random features of
the orbit of the ball, having of course nothing to do with the geological
features of the region being hit. However, the probability for kimberlites
produced long ago having been
preserved in such a way as to be {\em observed} today
is not guaranteed to be independent of the stability of the
area onto which the fall took place. Naturally a very stable intraplate
region may keep its kimberlites undisturbed much longer than a region
with much sedimentation activity. So statistically we expect to find more well
conserved kimberlites in such intraplate regions, as seems to be the case.

The fact that in the same region we can find clusters of different age
is of course expected in our model. In our picture the
different clusters should be of completely different age, since the
falls are unrelated to earthly happenings.
However, peaking of the time distribution for kimberlite formation in certain epochs
is {\em not} so welcome for our theory.

\subsection{On the distribution
of kimberlites}

It is our hypothesis that the phenomenon
known as kimberlites is due to our
dark matter balls. That is to say we
imagine that each time a ball falls
on the earth
the ball digs a long tube which
successively gets filled with material
that comes up from deeper layers,
perhaps from the mantle. Kimberlites
understood as such tubes have been
found in various places on the earth.
But they are very far from being
smoothly distributed all over the earth
surface, as one would naively expect
of a phenomenon coming from random
hits from the galaxy halo. But that
expectation is, of course, only for
the distribution that would be observed
provided one were able to observe
all the kimberlites, even if they had
been heavily covered by sediments.

In this section we  shall rather attempt to make a crude
estimate of how many fewer kimberlites are expected to be
practically accessible and thus may have been truly found. We
shall also remark that while sediments will quickly make a
kimberlite practically invisible, erosion will not prevent the
visibility of a kimberlite. This is because the tubes are supposed to go
so deep that erosion may in practice never go sufficiently deep so
as to reach the bottom point of the kimberlite (where in our point
of view one might find the dark matter ball situated). Rather erosion could
have eroded away some sediments, so that places where deep erosion
has taken place should be the most likely places to find the
kimberlites. This is indeed exactly the experience: The
kimberlites are mainly found in the Archean cratons, meaning in
geological provinces where there has indeed been mainly erosion so
that very old geological layers are on the surface.

\subsection{Statistical estimate of
numbers of kimberlites}

With the earth having existed for of the order of 4 milliard years
and one ball falling around every two hundred years, we would
naively expect to find of the order of $4*10^9/200 = 2*10^7$
kimberlite clusters.
With a surface area of the earth of the
order of $(10000\ \hbox{km})^2$ = $10^{14}\ \hbox{m}^2$, this would
mean a typical distance from one kimberlite cluster to the next of
the order of $\sqrt{10^{14}/(2*10^7)} \approx 2*10^3\ \hbox{m}$.
That is to say there should be only about a km from one cluster of
kimberlites to the next. But one has not at all found so many
kimberlites as this would mean! So either our hypothesis is
completely wrong or these many a priori expected kimberlites are
largely not observed at all.

Of course it is not unlikely that some
sediments have fallen on top of a
kimberlite site and covered it, so as
to make it unobservable.
We shall now make a very
primitive statistical model to estimate
the number of kimberlites which we
should really see.

\subsection{Introduction to statistical model}

The statistical model which we propose
describes the geological history for a single
site, meaning here just whether there is
sedimentation or erosion going on at the
site in question. To take definite
presumably reasonable numbers we assume
that for periods of the length of an
ice age, taken to be of the order of 10000 years,
one has either constant sedimentation or constant
erosion.
However whether we have
sedimentation or erosion varies
randomly from one period to the next.

Now the assumption is that we - the
geologists - only ``see'' (discover)
those kimberlites which are
at present
at sites more deeply  eroded than that site
has ever been eroded before. That is to say
we only ``see'' the kimberlites on sites
where all sediments ever settled have
been eroded away. Otherwise of
course the kimberlite would be covered
by sediments and we would most likely
not discover it.

On a given site the surface of the earth has been going up
and down - in our model as a random walk - due to
sedimentation and erosion respectively.
The probability distribution of the
depth into which there is total erosion
at a given place is of course a gaussian,
corresponding to the random walk of erosion
or sedimentation.
We take 2 milliard years as the typical time that has elapsed
since the fall of a dark matter ball responsible for a
Tunguska-like event. Then the typical random walk has
(2 milliard years)/(10000 years) = $2*10^5$ steps.

\subsection{Estimating probability for pipes being exposed}
Let us think of the sedimentation and erosion process at a site
as a random walk of, say, $n$ steps of the exposed surface up and down relative to
the material in the column under the site.
What we think of as a step is the activity occurring in our
period of 10000 years during which we assumed, for simplicity,
that erosion or sedimentation went on in only one direction.
The condition that
one can see the kimberlite pipes on a site is that they must be exposed,
meaning that any sediments that have fallen on the site - since
the fall of the dark matter ball - should have been eroded
away. In terms of the random walk this means that at the end time step
number $n_0$ after the fall, where the level is today, this $n_0$th step
shall be to the erosion side of the levels of all the previous steps.
In other words this last $n_0$th step shall be the hitherto deepest.
If we denote by $r$ the number of levels that this endpoint is
under the starting level when the ball fell, it means that the condition for
the kimberlite pipe being exposed is that it is the first time the
random walk reaches that depth, i.e. to $r$ or deeper. For a random
walk we denote the number of (time) steps to the first hit of a given
value $r$ as $T_r$.  If $S_n$ denotes the level reached after $n$ time steps
in a certain random walk then for this walk
$T_r$ denotes the number of time steps to the {\em first} case in which
$S_n = r$. In other words
\begin{equation}
 T_r = \hbox{min} \{ n\ge 1 \ \hbox{such that} \ S_n =r \}.
\end{equation}
In order that the kimberlite pipes at a site, in which there has
been erosion (into the material below) corresponding to a net
number of levels $r$ (i.e.~the number of time steps with erosion
minus the number of time steps with sedimentation), be exposed, it
is needed that in the random erosion-sedimentation walk this level
$r$ is hit for the first time - since the fall of the dark matter
particle. That is to say it is needed that $T_r = n_0$. Of course
we have then for the actual walk $S_{n_0} = r$. The conditional
probability, given the site,
 for the kimberlite pipes being visible is then\footnote{Equation \ref{tr}
is trivial in as far as we of course can write
the total (unconditional) probability $P(T_r =n_0)$ for the first hit
of level $r$ being at time step number $n_0$ as
\begin{equation*}
P(T_r = n_0) = P(S_{n_0} = r) P(T_r=n_0| S_{n_0} =r) + P(S_{n_0} \neq r)
P(T_r=n_0|S_{n_0} \neq r).
\end{equation*}
Now hitting $r$ the first time at $n_0$ of course means that it also hits
$r$ at time step $n_0$, so that $P(T_r = n_0|S_{n_0} \neq r) =0.$}
\begin{equation}
 P(T_r = n_0 | S_{n_0} = r)= \frac{P(T_r = n_0)}{P(S_{n_0} =r)}.\label{tr}
\end{equation}
Now, from the hitting time theorem \cite{rw}, we find the random
walk probability formula for this case to be
\begin{equation}
 P(T_r = n_0) = \frac{r}{n_0}* P(S_{n_0} = r)
\end{equation}
for $r\ge 1$. Thus
\begin{equation}
P(T_r=n_0|S_{n_0} =r) = \frac{r}{n_0}.
\end{equation}

It is well-known that the distribution of $S_{n_0}$
is approximately a Gaussian with a width growing
as the square root of $n_0$.  So it is very unlikely in our model
to find a site where $r$ is more than a few times $\sqrt{n_0}$.
This in turn means that we can count the order of magnitude
of $\frac{r}{n_0}$ as being typically $\approx \sqrt{\frac{1}{n_0}}$.
While it is clear that in a site with more steps of sedimentation
than of erosion (i.e. $r < 0$) you can find no kimberlite pipes, the
probability for finding the possibility of kimberlite pipes being
exposed at a typical site (i.e. $r\approx \sqrt{n_0}$) is
\begin{equation}
 P(\hbox{exposed pipe}) =P(T_r = r |S_{n_0} = r) = \frac{r}{n_0}
 \approx \frac{1}{\sqrt{n_0}} \approx
\sqrt{\frac{10000\ \hbox{years}}{2*10^9\ \hbox{years}}}
\approx 2.2 *10^{-3}.
\end{equation}

This means that on a large scale it is only about $2.2 *10^{-3}$
of the area in which the kimberlite pipes will be visible. So,
even in say the Canadian Shield\footnote{The Canadian Shield is a
broad region of pre-Cambrian rock that encircles Hudson bay.},
the number of kimberlite pipes
expected to be detectable is this number $2.2 *10^{-3}$
multiplied by the number of dark matter particles falling on that
region.  Now the area of this Canadian Shield is about $5.5* 10^6 \
\hbox{km}^2$, while the total earth surface area is $500 *10^6\
\hbox{km}^2$. So the Canadian Shield makes up $\approx 1 \% $ of
the earth's surface and, with one dark matter ball falling every
200 years for 2 milliard years giving $10^7$ balls on the whole
earth, we expect $10^5$ balls to fall on the Canadian Shield. So
the number of clusters of kimberlite pipes expected to be exposed
on the Canadian Shield becomes
\begin{equation}
 \hbox{\# \hbox{exposed pipes}} \approx 2.2*10^{-3} * 10^5 \approx 200.
 \label{exposed}
\end{equation}
This is to be compared with the at least 30 observed clusters
\cite{Barbara}. However if you count all the pipes rather than the
number of clusters, the number observed in Canada is more than
770. That is to say the estimate from our model predicts about 6
times as many kimberlite pipe clusters as found, if you imagine
that each Tunguska-like event would produce a whole cluster. But
if you imagine that each Tunguska-like event only produces one
pipe, we rather predicted a factor 4 too few. So using the
ambiguity in what to count in order to obtain the number of
Tunguska-like events as an uncertainty, our prediction agrees with
the observed number within errors.  But now our estimate is indeed
very uncertain in itself and we shall make an estimate of this
uncertainty in the next subsection.

\subsection{Uncertainty estimate for number of kimberlite pipes/clusters}
The most uncertain input to our estimate of the number of exposed
pipes is presumably the time scale for the typical alternation of
sedimentation versus erosion, which we took to be $\Delta t =
10000 \ \hbox{years}$, just referring to an ice-age. This 10000
years number could easily within uncertainty be a factor up to a
100 times bigger or smaller. So we take the logarithmic
uncertainty in the typical time scale to be
\begin{equation}
\Delta \ln(\Delta t) \approx \ln(100) \approx  4.6.
\end{equation}
Let us now similarly estimate the uncertainties of the other input
quantities:

The time since the fall of a dark matter ball should at least be shorter than
the age of the earth, which is
supposedly 4.6 milliard years. We took 2 milliard years as the central value
in our estimate above.
Now on the other hand the times of the creation of the kimberlites
in South Africa seem
\cite{Jelsma} to be dominantly less than about 200 million years ago,
although a few are indeed older. Thus effectively the upper limit
for the age of the detected kimberlite pipes lies in the interval
from $2*10^8$ years to $4*10^9$ years. We describe this as
$2*10^9$ years with a logarithmic uncertainty
\begin{equation}
 \Delta \ln(t_0) \approx \ln(\sqrt{20}) \approx  1.5.
\end{equation}

The rate $r_B$ of Tunguska-like particle falls is of course very
uncertain, in as far as we have so far only seen {\em one}
Tunguska event. We should therefore at least let our estimate of
200 years for the time between Tunguska-like events be uncertain
with a factor $e=2.718$.
 So we take\footnote{We include the uncertainty in what should be
 counted as the
 earth surface area (i.e.~whether only the land or also the sea
 area should be counted)
 in this value for $\Delta r_B$.}
\begin{equation}
 \Delta \ln(r_B) \approx 1.
\end{equation}

The uncertainty in the area of the Canadian Shield may not be great,
but why should some precise definition of the Canadian Shield
be exactly where the kimberlite pipes remain visible? So we may take
this area also to have an uncertainty of say a factor of about 2 to be
realistic
\begin{equation}
\Delta \ln (\hbox{area of Canadian Shield}) \approx 0.5.
\end{equation}

Then we shall remember that, in the calculation we did
above, the end result for the number of exposed kimberlite pipes (\ref{exposed})
expected in the Canadian Shield depended on $t_0$ and $\Delta t$
only through their square roots, while $r_B$ and the area of the
Canadian Shield came in as simple factors. Thus the logarithmic
uncertainty in the final number of 200 exposed pipes obtained in
the last section becomes
\begin{equation}
 \Delta \ln(\hbox{\# exposed pipes})
 \approx \sqrt{(4.6/2)^2 + (1.5/2)^2 + 1^2 +0.5^2} = 2.7.
\end{equation}

So the result of our prediction for the number of kimberlite pipes
expected to be found exposed on the Canadian Shield is
\begin{equation}
 \hbox{\# exposed pipes} = 200 *\exp(\pm 2.7) = 14 \
\hbox{to} \ 3000.
\label{uncertainty}
\end{equation}
This result should be compared with the 30 or so observed clusters\footnote{In some cases
these so-called clusters contain pipes with different ages and thus should certainly be
counted as more than one event in our picture; so perhaps we should increase this number
to 40 or more "independent" clusters.} and 800 or so individual pipes. These two numbers,
30 and 800, {\em both} lie inside the uncertainty range (\ref{uncertainty}) of our
theoretical prediction.

So our hypothesis that Tunguska-like events are the reason for the formation of
kimberlite pipes is at least not numerically excluded order of magnitudewise
with respect to the number of pipes.
However a major problem for our model is that kimberlites are
distributed in a quite structured way in time \cite{Jelsma, Heaman} with the weight
on the rather young kimberlite pipes being detected.
In our model of random walk sedimentation versus erosion clumping of
time intervals in which kimberlite pipes survive to be exposed today is also possible,
but really the distribution found is for us
not the obvious one expected.

\section{Penetration depth}
\label{penetration}
We shall here first estimate how deep our ball will penetrate into the
earth using essentially dimensional arguments. Then we shall
attempt to gradually improve this estimate.

A) The first very crude argument is the following:

The material in front of the ball as it passes through the earth has somehow
or another to be pushed to the side, so that the ball can pass by without
having the earth material penetrate into the ball (we assume that the ball
is so dense and so solid that no earth material can penetrate into it).
In the crude approximation of treating only the main part of the earth
in front of the ball, the velocity needed for it to escape going into the
ball is of order of the ball velocity. The slope of the side of the
ball relative to the direction of motion of the ball corresponds to an angle
of order unity, and thus the escape velocity for the material has to be of the same order
as the ball velocity.
Let us at first assume that the
amount of material speeded up to this velocity
is only a factor of order unity bigger than the absolutely necessary amount
lying in the way  of the ball. Then, order of magnitudewise, the
kinetic energy that has to be provided to push the material away is of the
order of the the energy needed to give the column of material in
the way of the ball the speed of the ball.
Now suppose all the kinetic energy of the ball is transferred
to the pushed away material in this way.
This would mean that the mass of the column
of material in the way of the ball should be, within our accuracy,
of order unity times the mass of the ball.
Taking the
specific weight of the earth material in the region penetrated by the ball
to be $\rho_{earth} = 6000\ \hbox{kg/m}^3$,
the penetration depth $l$ is given in the first crude approximation by
\be
l_0 = \frac{m_B}{\pi R^2 \rho_{earth}} = \frac{ 1.4 * 10^8 \ \hbox{kg}}{\pi *
(0.67 \ \hbox{cm})^2 * 6000 \ \hbox{kg/m}^3}
=1.7 *10^5\ \hbox{km}.
\label{l0}
\ee

We believe this first estimate $l_0$ appreciably overestimates the true
penetration depth $l$, as we shall argue below where we present a series
of correction factors.

B) The material that has to go around the ball and not be carried along with it
must use at least some region of the cross section outside the ball itself.
Thus a larger column of material has to be speeded up to the speed of the order
of that of the ball. If, for instance, this column corresponded to a doubling of the
diameter of the disturbed region,
then the area $\pi R^2$
in our calculation would have to be increased by about a factor 4.
Subsequently the penetration depth $l_0$ would be decreased by a factor 4.

C) Also the material cannot truly escape just with exactly the same speed
as that of the ball itself, because the closer one gets to the very front of
the ball the faster must the material move in order to escape.
However, we may imagine that, under the motion, the ball could have a cone of
very strongly compressed material in front of it, which could push the
material being met away as the ball progresses. Thereby the angle of the effective
surface of the ball with the direction of motion could be prevented from
being just $90^{\circ}$ as it a priori is just in front of the ball.
Such
a $90^{\circ}$ angle between the direction of motion and the direction along the
effective ball surface would mean that the velocity needed to escape at such a
place would be infinite.
This fact indeed suggests that there should be
some peaked material in front of the ball, so as to avoid such an infinite speed.

Let $\theta$ denote the spherical polar angle for the points on the surface
of the ball with $\theta = 0$ at the front of the ball in its motion with
velocity $v$. Then the escape speed needed by the earth material at the surface of the
ball in the direction transverse to the direction of motion of the ball is
$v_{tr} = v\cot\theta$.
Then we can formally evaluate the average of this transverse speed squared
over the surface of the ball (which we need to estimate the kinetic energy of
the earth material) as the integral form
\be
<v_{tr}^2> = v^2\int_{0}^{\pi/2} \cot^2 \theta  \sin\theta \, \mathrm{d}\theta=
 v^2\int_{0}^{\pi/2}  \frac{\cos^2 \theta}{\sin \theta}\,
\mathrm{d}\theta.   \label{integral}
\ee
This integral is divergent at the tip of the ball where $\theta = 0$.
It diverges logarithmically. The typical (non-divergent) part of the integral
is of order unity.
Presumably the logarithmic divergence will be cut off
due to a nose of compressed
material sitting at the front of the
ball, giving the ball an effective shape with a peak at the front.
This will modify the integral (\ref{integral})  near the
front point of the ball and so remove the logarithmic divergence.
We so to speak obtain the logarithm of the nose-angle.

We guess that this logarithm may be of the order of 3 or 4, giving
$<v_{tr}^2> \sim 4v^2$. Hence our estimate of the kinetic energy
deposited in the earth per unit length is increased by a factor of, say, 4.
This corresponds to reducing the penetration depth by another factor of 4.

D) We may think of the motion of the material pushed out from the pathway of the
ball either as a macroscopic motion of material flowing as a whole, or as
ions/particles moving essentially in a thermal motion ensuring the escape
of the particles in front of the ball.
While the material in front of the ball gets heated up in this way
to a sufficiently high temperature that it has the velocity needed to escape,
the ball itself must actually still
keep to a much lower temperature.
So during the main part of the ball's motion we must consider the ball itself
to be cold compared to the temperature of the material in front of it. Thus
a fraction of the energy of the material pushed away from the ball
will be lost by heat conduction into the ball, heating the latter up.
Let us say that $K$ times the energy deposited in this  material as heat
due to its interaction with the ball is conducted into the ball.
Then it means that the amount of energy per unit time needed to
get the material pushed out with a prescribed speed - to avoid
being in place where the ball comes by - gets increased by a
factor $1+K$, simply because for each joule deposited $K$ joules
stream into the ball heating it up. But this extra energy
proportional to $K$ has to come from the work done by the ball as
it presses its way into the earth. Thus the estimated braking
force acting on the ball must be made stronger by the factor
$1+K$. This means that the estimated stopping distance of the ball
must be shorter and that the penetration depth should be reduced
by a factor of $1+K$.

So it becomes important to estimate if possibly
$K$ could be appreciably larger than $1$.
When say a nucleus hits the ball, it may effectively hit a few
electrons or nuclei in the ball and, after such a collision, it
would still have an energy that well may have decreased but not by
a dramatically big factor. Now, however, the same say Si-nucleus
in front of the ball could hit the ball many times and thus could
lose a much bigger fraction of its energy than by just hitting
once. However, in this case, the nucleus would have hit another
particle somewhat further away from the ball, which may
consequently not hit the ball. So, on the average, we expect the
particles in front of the ball only hit the ball of order once
before they escape from the path of the ball.
Thus we conclude that $K$ should be of the order of unity, although it
could be somewhat bigger than 1. For definiteness we shall take $K=2$,
leading to a suppression of the penetration depth by a factor of $1+K=3$.

   E) So far we have only really counted the kinetic energy and have ignored
the fact that some energy will be deposited in the earth as potential energy.
As a rough approximation we can take the potential energy deposited to be the
same as the kinetic energy deposited in the earth. This has the effect of
reducing the penetration depth by a further factor of 2.

As a conclusion of this section we may claim:

Most crudely we got that the mass of the material
penetrated in a column with the cross section of the ball
should be just the mass of the ball. This gave the first
estimate (\ref{l0}) of the penetration depth to be $l_0 = 1.7*10^5$ km.
However there were several ways of making a somewhat more realistic
estimate. All the proposed corrections were in the direction of
increasing the energy needed to push material aside by the ball.
Thus a bigger energy loss per unit length by the ball is needed.
In fact the above four corrections B)-E) lead to an increase in
the energy loss and hence a decrease in the penetration depth $l$ by a
factor of order $4*4*3*2 \sim 100$. Thus we finally end up with a
penetration depth of
\be
l= l_0/100 = 1700 \ \hbox{km}.
\ee
as our best estimate. However it is of course extremely crude and could be
wrong by an order of magnitude or more.

A priori this penetration depth means that the ball would penetrate deeply into the
earth and remain buried there.
However, within the accuracy of our estimation, it could nonetheless be the case that
the ball would pass straight through the earth. In any case the total ball energy
or a large fraction of it would be deposited in the earth.
Then this energy
would have either to stay in the earth or come out by some tubes extending from its
route through the earth or simply backward along the way the ball came in.

\section{How were the balls formed in the
early universe?}
\label{early-universe}

If, as we assume from the Multiple Point Principle, the vacuum
{\em with} and the vacuum {\em without} the bound state condensate
have the same energy density, there will at temperatures higher
than the weak scale be approximately equal amounts of the two
types of vacua. In fact a high density of the walls separating the
two vacua will be present due to thermal fluctuations around in
space. The amounts of space covered by each of the two vacua will
be about the same.

As the temperature $T$ falls with cosmological time $t$ according to the order of
magnitude formula\footnote{Here we use a reduced Planck mass, which we take to be
given by $M_{Pl} = 1\ \hbox{s}\ \hbox{MeV}^2$ for the purposes of this section.}
\begin{equation}
t * T^2\approx  1\ \hbox{s}\ \hbox{MeV}^2\approx M_{Pl},\label{tT2}
\end{equation}
the walls tend to contract and there are less extensive
walls. Consequently when the temperature falls below the weak scale (say 100 GeV)
the local extensions of the phases/vacua become larger. The balance
between the two phases/vacua becomes unstable, in the sense that
due to some possibly small effect one of the phases has become a bit
less copious than the other one and will tend to contract further.
In the appendix \ref{whichvacuum}, we argue that it is the phase with a
condensate which contracts.

At first one would estimate that the walls would quickly reach
velocities of the order of the speed of light for dimensional
reasons. However,
at several temperatures the walls tend to brake
the passage of one species of particle or the other. Consequently the
plasma of the mixture of the various particles tends to be driven
along with the walls.

If indeed the coupling of the walls to the
plasma is sufficiently strong, one could have imagined that  the temperature and
pressure inside
a vacuum region about to contract will rise and soon slow down the
contraction (relative to the general Hubble expansion\footnote{In
this high temperature era even the balls which are about to
contract will in reality typically expand due to the Hubble
expansion.}) until heat can  be conducted out and thus again allow
the contraction.
However, it turns out that neutrinos can transfer heat so efficiently
that the balls cannot attain sufficiently high temperatures inside to
prevent collapse due to such an effect.

There is, however, an effect causing a higher inner pressure in the with condensate balls,
so as to drive them towards growing rather than contracting.
At different temperatures this effect originates
from those Standard Model particles which have their mass scale in the range around the
prevailing temperature. Basically the effect is that, due to the lower Higgs
expectation value in the with condensate phase, the masses of the particles obtaining
their mass from the Higgs field become smaller in the with condensate phase than in the without
condensate phase. This in turn means that, at the various temperatures, one or the other of the
particles has a higher particle density in the Planck plasma in the with condensate phase than in the
without condensate phase. This pressure then acts to expand the balls consisting of the with condensate phase.
This effect means that sufficiently big balls would expand further and not contract at
all. We refer the reader to appendix \ref{survival} for more details.

\subsection{Discussion of survival of the balls}
The crucial point for our model to work producing dark matter in
the early universe is that the balls of the required size should
survive - i.e. avoid collapsing - long enough that they can be
stabilised by being filled with nucleons, and thus survive
forever. Since we estimated in section \ref{qmd} that the binding
energy of a nucleon into the phase with the condensate is of the
order of $10\ \hbox{MeV}$, the concentration of the nucleons into
the phase with the condensate will only be truly active when the
temperature has fallen to $T =10\ \hbox{MeV}$.

In appendix \ref{survival} we estimate that the smaller masses of several
Standard Model particles inside the balls, where the Higgs field VEV and
thus these masses are smaller, cause an increased pressure from these
particles tending to pump up the balls. This effect is estimated in
(\ref{Rcrit}) to be so strong in the $T = 10$ MeV era as to make balls
with radius $R$ greater than $R_{crit} \approx 2 \ \hbox{mm}$ grow
rather than contract. These balls can avoid collapse until the temperature
is suficiently low that nucleons start to collect inside them.

\subsection{The phenomenologically required balls}
In subsection \ref{sec:ball-size} we have estimated that the balls
we want to form should at the end get a size of around 0.67 cm.
These balls should have quickly collected nucleons once the
temperature fell sufficiently low for them being able to catch
them - namely below about 10 MeV.
In fact we estimate, in appendix \ref{fadingout}, that the balls in this
10 MeV era run around in a rather complicated motion and even hit each other,
thereby sweeping up essentially all of the nucleons.
Remember that we imagine that at the weak
scale of temperature the two vacua filled comparable amounts of
the total volume, so that the distance between the ``balls'' at
that time was of the order of their typical size. This size should
have then crudely followed the Hubble expansion.  So the size, before the
partial collapse to their final stable size,
should be of the order of the  distance between the balls. We
earlier estimated, see equation (\ref{lT}), that the distance
$l_T$ between the phenomenologically wanted balls at the 10 MeV
era was of the order of 6 m. So, for our model to work, we need
that this wanted size of 6 m has to be bigger than the critical
size below which the balls would have collapsed before they get,
so to speak, rescued from total collapse by the nucleons piling up
inside them. In the foregoing subsection and appendix \ref{survival}
we saw that the critical
size for survival to the 10 MeV era was 2 mm.
Since 6 m is appreciably bigger than 2 mm it looks, even though we
have made only crude estimates, very likely that the balls, which
we postulate constitute dark matter, could have survived until the
era when nucleons can fill them up and prevent their total
collapse.

\subsection{Capturing the nucleons}
\label{capturing}
Once the temperature falls down to around $T \approx \Delta V
\approx 10 \ \hbox{MeV}$, the nucleons start to collect into
the phase {\em with} a bound state condensate. This phase is
then favoured by the Boltzmann distribution for the nucleons,
because the Higgs field and thereby the quark masses are
smaller there. However, in appendix \ref{heatconduction}, we have estimated
the rate of diffusion of nucleons in the plasma to be too
slow for them to spread throughout the volumes of {\em with}
condensate phase in this era. Rather it appears that, if the walls
between the phases do not move around significantly, there will only
be a pile-up of nucleons in a thin layer  on the {\em with} condensate side of
the walls. This pile-up would be formed by pulling over nucleons from
a thin layer on the {\em without} condensate side of the wall.
In this case many of the nucleons would remain outside the balls in
the phase without a condensate.

However, in appendix \ref{fadingout}, we consider the motion of the walls
and find that they move rather freely at least when the temperature has fallen
below 20 MeV. Equation (\ref{taut}) shows that, for
wavelengths characteristic of typical balls,
the time for the waves to die out - the  survival time - is larger
than the Hubble time for temperatures below 20 MeV.

For example  consider the situation for a typical ball which ends up as dark matter;
at a temperature of 10 MeV it has a radius of order 6 m. So we now use equation
(\ref{taui2}) to evaluate the
survival time $\tau_i$ for the following parameters: ``wavelength" = 6 m, $T = 10\ \hbox{MeV}$,
$S^{1/3}= 28 \ \hbox{GeV}$.
\begin{equation}
\tau_i \approx g^4 \left (\frac{28\ \hbox{GeV}}{10\ \hbox{MeV}}
\right )^3 *6\ \hbox{m}
\approx  \left \{ \begin{tabular}{cc c } 400 \ \hbox{s}& for & $g^2=1$\\0.04\ \hbox{s} & for & $g^2=1/100$.
                            \end{tabular}\right .
\end{equation}
This means that the waves of this relevant length first decay after a time interval lying
in the range between 0.04 s and 400 s.
However that is longer than the Hubble time at the 10 MeV era, which is only 0.01 s.
This means that, in the 0.01 s era, the walls run with almost negligible friction through
the plasma. They are essentially freely moving walls and their inertia becomes very important.


The walls will thus clash together and at least sometimes get thrown away from each other again
so as to run on.
During such processes new regions of the with condensate phase can get created.
Nucleons will be caught in such regions formed by the running walls, provided the
temperature has fallen to less than or about the $\Delta V = 10$ MeV value.
Because almost all space will soon have
been passed through by these running walls, it becomes likely that all the nucleons in the
Universe quickly get collected into the with condensate phase once the temperature
fell to the order of
$\Delta V = 10 $ MeV. Thus we expect that before the contraction of the balls,
at a temperature of the order of 2.3 MeV
(see equation  (\ref{contraction}) in section \ref{bbn} below),
all the nucleons will already have gotten collected into the with condensate phase.

\subsection{Ordinary versus Dark Matter}

Depending on the details of the development of the dark matter
balls, they will collect a bigger or smaller fraction of the
excess nucleons in the plasma. If there happens to be an
era after the temperature has gone below ~$\Delta V \approx 10$
MeV in which there is a relatively free passage of nucleons in and
out of the balls, then the Boltzmann distribution of the nucleons
will mean that almost all the nucleons get into the dark matter
balls. Really we argued above in subsection \ref{capturing}
that this is indeed the case.

So practically all the nucleons go into the balls which
eventually come to make up the dark matter. If the transport of
nucleons across the skin of the balls stopped at this time, there
would be essentially no ordinary matter. However it is quite
possible that the dark matter balls can expel some of their
nucleons and thereby supply some ordinary matter. Indeed we have
earlier published \cite{crypto} the idea that nucleosynthesis is
likely to have occurred inside the balls during the late stages of
their contraction. Emission of nucleons could have been the main
way of getting rid of the excess energy, released due to the
increased binding energy of the nuclei, in some of the steps in
this process. In fact we suggested that this cooling by nucleon
emission dominantly occurred during the transition from helium to
all the heavier nuclei. It should then be understood that the
formation of helium occurred at a lower nucleon density with the
heat dissipated rather by the emission of gamma rays and/or
electrons and positrons. We briefly discuss these nuclear physics
processes in appendix \ref{nuclearphysics}.

Actually the main point of our previous work was to call attention
to the following  numerical coincidence:

The increase in binding energy in going from helium to essentially
all the heavier elements is equal to 1.4 MeV per nucleon. Now
suppose that the ball would irradiate this amount of gained
energy, by sending out nucleons with a small kinetic energy. Then,
for each nucleon emitted, the system would have to provide its
binding energy of 7.1 MeV in helium. The binding energy in heavier nuclei
is 8.5 MeV per nucleon\footnote{Here we use the nuclear binding energies
as measured in the normal phase, without a bound state condensate, and
assume that the ratio of the binding energies per nucleon between two
different nuclei is the same in both phases of the vacuum.}.
So for each nucleon emitted from a ball on the borderline of stability, there should
be $8.5/1.4 \approx 6$ nucleons becoming bound into heavy nuclei.
It follows that if, under the high nucleon density inside the
ball, all the nucleons remaining inside formed heavy nuclei then a
fraction of about $1/6$ should be emitted. It is these emitted nucleons that should
be identified with the ordinary matter and those remaining inside
the balls identified as dark matter. The ratio of ordinary matter to dark
matter is then predicted to be about $1/5$, in good agreement with
astronomical data.
In appendix B.5 we repeat the calculation for median size balls
and obtain a prediction for this ratio of ordinary to dark
matter of
1/5.6, which is to be compared with the recent Planck measurement
\cite{planck} of 1/5.44.

It is very important that such a process of creating ordinary
matter, or rather separating it from the dark matter, takes place
before the Big Bang Nucleosynthesis gets under way.
Otherwise it could disturb the usually so successful understanding
and fitting \cite{pdg,Olive} of the abundances of the light elements by Big Bang
Nucleosynthesis. Especially the nucleons emitted by our balls
would tend to change the neutron-proton ratio, unless the weak
interactions were still active in maintaining the Boltzmann
equilibrium ratio. So the separation of ordinary matter from dark
matter in our model must be essentially completed before the
neutrino decoupling temperature of 0.8 MeV is reached; otherwise
there is a severe danger that our model would change the
predictions from Big Bang Nucleosynthesis.

\subsection{Can the balls form before the Big Bang Nucleosynthesis?}
\label{bbn}
When do now these balls, which made up the ones we ``see'' as dark
matter, contract around their nucleons? The size of a ball still
following the Hubble expansion will be inversely proportional to
the temperature in the radiation dominated era. So the
phenomenologically wanted balls would have the size $l_T = (6\
\hbox{m}*10\ \hbox{MeV})/T$ at temperature $T$. When this size
becomes equal to the critical radius $R_{crit}(T)$ estimated in
equation (\ref{Rcrit}) of appendix \ref{survival},  these balls will collapse until
stopped by the nucleons inside them. This gives us the following
equation for the determination of the temperature era of the partial
collapse:
\begin{equation}
 6\ \hbox{m} *\frac{10\ \hbox{MeV}}{T} \approx  \frac{4 * (28\ \hbox{GeV})^3}{T^4}.
 \label{lTlcrit}
\end{equation}
However, as explained in eq.~(\ref{f152}) of appendix \ref{neff},
this order of magnitude formula
should be corrected as follows  with a few factors of order unity
\begin{equation}
 (1/2)^{1/3} * 6\ \hbox{m} *\frac{10\ \hbox{MeV}}{T} \approx
\frac{3 * (28\ \hbox{GeV})^3}{n_{eff}\sigma T^4}.
\end{equation}
Here $n_{eff} $ is the effective number of particle species times
their number of spin-components weighted with their efficiency of
interaction with the phase border and $\sigma=\pi^2/60$ is the
Stefan-Boltzmann constant;
the factor $(1/2)^{1/3}$ corrects for the fact that when the balls are
big - i.e. before their collapse - they only fill half the spatial volume.
This  gives the result that the temperature, in the era when the typical ball
considered contracts, is
\begin{equation}
 T_{contraction} \approx \frac{2^{1/9}*3^{1/3}* 28\ \hbox{GeV}}
 {(n_{eff}\sigma)^{1/3} * (60\
 \hbox{MeV m})^{1/3}}
\approx  \frac{0.65\ \hbox{MeV}}{(n_{eff}  \sigma)^{1/3}}
\approx 2.3\ \hbox{MeV}.\label{contraction}
\end{equation}
Here we used the value for $n_{eff}$ as estimated in eqs.~(\ref{f146}, \ref{f149})
of appendix \ref{neff}:
\begin{equation}
 n_{eff} \approx 4 *0.159 *\frac{7}{8} \frac{``\Delta m''}{T}\ \
 \hbox{for T around a few MeV,}
\end{equation}
where only the electron and positron are important.
In this formula we use the notation $(``\Delta m'')^2 = m^2_{without\ condensate} -
m^2_{with\ condensate}$ for the shift in mass squared between the
two phases of a particle species relevant for calculating the
pressure at the temperature $T$; the factor of 4  denotes the
number of particle plus antiparticle components for this species
(the electron in this case).
In practice we take $\frac{``\Delta m''}{T} \approx \frac{1}{4}$.

This temperature (\ref{contraction}) is to be compared to the
temperature at which the weak transformations between protons and
neutrons stop \cite{pdg,Olive} $T_f = 0.8\ \hbox{MeV}$  (the
freeze-out temperature). Now there is the little caveat that the
balls, which form, have of course some mass or size distribution
as discussed in appendix \ref{distribution}. However there is an
effective cut-off size given by the Hubble distance \cite{Kibble},
which at the weak scale temperature $T \approx 100 \ \hbox{GeV}$ is $1/H
\approx 1\ \hbox{cm}$. Such maximal size balls will collapse onto
their nucleons at the lower temperature
\begin{equation}
T_{contraction, maximal} \approx
\frac{3^{1/3} * 28\ \hbox{GeV}}{(n_{eff}\sigma)^{1/3} *
(100\ \hbox{GeV cm})^{1/3}}
 \approx  0.6\ \hbox{MeV}.
\end{equation}
Here we have used  $\frac{``\Delta m''}{T} \approx \frac{2}{3}$.

It follows that, for the majority of the balls after their collapse
to their final size of today, there should be
sufficient time in which the neutron proton ratio for the nucleons
expelled from the balls can take on its
thermodynamic equilibrium value. Even for the maximal size
balls it is probably also approximately true, since the
neutron freeze-out temperature
$0.8\ \hbox{MeV}$ is of the same order as the collapse temperature
for the biggest balls. Thus our balls, which are by then very
small in volume, will not disturb the development of the expelled
nucleons from behaving just as under the usual Big Bang
Nucleosynthesis calculations \cite{pdg,Olive}.
Since these calculations are very
successful in predicting ratios of light isotopes in metal poor
regions of the universe, it is important that we retain these
predictions in our model without having to refit with extra
parameters. It should be borne in mind though that of course the
very important $\eta$-parameter giving the baryon number relative
to the photon number used in the Big Bang Nucleosynthesis
calculations has to be carefully identified in our model; this baryon
number should be taken to be the number of baryons {\em expelled from the balls
under the fusion explosion}. That is to say that the  baryons
staying back inside the balls are {\em not counted into $\eta$},
although they are of course truly baryons and there are
appreciably more of them than the ones expelled. The crux of the
matter is that these baryons staying inside the balls are hiding behind each
other, so that at most the ones sitting on the surface of the balls
can at all be hit by the external baryons. In other words the vast
majority of the matter in the balls is effectively completely
inaccessible to the external nucleons and other particles. Compared
to the enormous weight of each ball and thereby its gravitational
interaction the balls are extremely ``dark'', in as much as they
almost do not interact in other ways than gravitationally.

\section{Conclusion}

We have put forward the idea that the Tunguska-event was indeed
caused by the impact of one of our dark matter balls. Such a ball
consists of a small piece of a new type of vacuum, characterized by having
a Bose condensate of our proposed bound state
of $6 t +6\overline{t}$ in it, and is surrounded by a skin separating the
new vacuum from the normal vacuum. Inside it is filled with ordinary
matter under a high pressure (caused by the skin or wall surrounding the ball).

The picture is supported by a few remarkable agreements:

\begin{itemize}
 \item { Impact rate leads to weak scale tension in ball wall}

The rate of in-falls of the proposed balls onto the earth is
exceedingly crudely estimated to be $r_B\approx 1/(200 \
\hbox{years})$, just from the fact that we saw only one ball about
one hundred years ago. Further we assume that the balls are typically
of such a size that they only {\em barely avoid collapsing} by
getting the nucleons pressed out through the wall. These two
natural assumptions lead to the very nice agreement with the idea
that the tension $S$ of the wall is given by the weak interaction
scale $S^{1/3} \approx 100$ GeV. (Really our best estimate in
appendix \ref{solitonic} for
the condensate wall tension is $(16 \ \hbox{GeV})^3$, while
our estimate using the presumed rate
of Tunguska falls and critical ball size gives
the value $(28 \ \hbox{GeV})^3$
for the tension (\ref{e42}).)

\item {Explosion energy is comfortably smaller than the estimated kinetic
energy of the ball}

The Tunguska explosion energy has been estimated to be about 10 to
30 megaton TNT. On the other hand, the mass we got from the
density of dark matter, the impact rate $r_B$ and the typical halo
velocity gives a kinetic energy for the ball corresponding to 430
megatons TNT. But this kinetic energy is bigger than the Tunguska
explosion energy by a factor 20.  Indeed it  is quite likely that
1/20th of the energy of the ball would have appeared on the
surface of the earth in a short time after the impact.

\item{Dark to Normal Matter}

We have previously speculated \cite{crypto} that at some
moment - when the temperature of the universe was about 2.3 MeV
according to (\ref{contraction}) - an explosive fusion of helium
to heavier elements took place inside our balls. This would have
happened at a time when the typical balls had already contracted so
much that explosively expelled nucleons would no longer be captured
by other balls, because by then the balls would have already made up a too
tiny part of the volume. Simple energy estimates suggest, from the
increase in binding energy from 7.1 MeV per nucleon in helium to
8.5 MeV in the heavier elements, that about one
nucleon out of  6
nucleons were expelled. The expelled nucleons should become the
ordinary matter, and thus we explained \cite{crypto} the ratio of ordinary to
dark matter to be about 1/5 in good agreement with astronomical
fits. In appendix B.5 we have repeated the calculation for median size balls
and found this ratio of ordinary to dark matter predicted to be
1/5.6,
while the recent Planck data \cite{planck} are fitted by the value 1/5.44
for this ratio.

\end{itemize}

Just the idea that the Tunguska event was caused by some small object
of extremely high specific density could be helpful in solving
some of the mysteries of the Tunguska event. Such an object
would penetrate so deep into earth that nothing from the object
itself  would be seen on the surface. Most obviously such
a rather small but, for its size, extremely heavy particle would
not be mainly observed via its track in the atmosphere but rather by
hot earth material being seemingly thrown out from the earth.
This material would presumably emerge from funnels, much like
kimberlite pipes, such as the Suslov Crater, Lake Cheko and perhaps
the funnels observed by Kulik \cite{Kulik}.
If such small objects of high density were
moving at a relativistic speed, they might not be observable
astronomically and cosmologically today as dark matter since the latter should be
``cold''. But suppose they are
non-relativistic and with a typical, say, halo speed of the order of
160 km/s as discussed in the present article. Then if such a
particle  should deposit in the air above Tunguska\footnote{Here
we also assume such particles cause other Tunguska-like events
at a rate of approximately once every 200 years} about 1/20 of
the energy deposited at all in the earth, the total amount of
energy and thereby also of mass carried by these particles
would have
to be like that of the dark matter in the halo. So, under the
assumption that the particles are in the halo of our galaxy, they
would have a total mass similar to that of the dark matter there.
Thus it would be hard not to identify them with the dark matter; otherwise there
would no longer be place for the dark matter. Of course for such a
particle of the non-relativistic type to deliver the energy suggested
by the Tunguska event, we need also the particle mass to be about the one in
our fit, namely about $10^8$ kg.

It is in general a bit of a mystery why the density of dark matter in the
universe is so close to the density of ordinary matter, only deviating
from it by a factor of 5.
Supersymmetry models with WIMPs can at best obtain this result as being a bit
of an accident. A model like ours, in which the dark matter at the
root of it consists of ordinary matter bound into small objects,
should a priori have a better chance to cope with this mystery.
As already mentioned we even get just the right ratio of
dark to normal matter. But then the needed high specific density
requires a very strong compression or some alternative mechanism
for obtaining an anomalously high specific density. It is also
crucial that the part of the ordinary matter presented as dark
matter gets so strongly compressed that it is effectively
completely inaccessible under the Big Bang Nucleosynthesis;
otherwise the good agreement with data of the Big Bang
Nucleosynthesis calculations \cite{pdg,Olive} with only
gravitationally interacting dark matter would be spoilt.

In this article we presented a picture of the balls being
formed, by having two types of vacua with somewhat different Higgs
field expectation values in the two vacua. These VEVs are assumed
to deviate by a factor of order unity, with the Higgs field
being smaller in the vacuum inside the dark matter balls.

Let us immediately stress the very important property of this model:
{\em the model
only uses the Standard Model,} so that no new physics is to be
assumed. This is contrary to the majority of models for dark
matter in high energy physics, which typically need for
instance supersymmetry so as to get supersymmetric partners
functioning as dark matter particles. It should though be admitted
that we make an extra assumption. This assumption does not truly
modify the Standard Model, but only provides a way of fixing
some of the coupling constants. In fact we assume the so-called
``Multiple Point Principle'' (MPP), which postulates that there should be several vacua
having the same energy density. We note that the MPP prediction
\cite{tophiggs} for the
Higgs mass is in remarkably good agreement with the recent LHC
measurements.

For the present article the crucial assumption is that there exists
one vacuum with a condensate of bound states of $6t +
6\overline{t}$ and one without this condensate; furthermore {\em
these two vacua have the same energy density}.
The existence of such a condensate requires a fine-tuning of the
top quark Yukawa running coupling constant $g_t$ \cite{nbs}. We have made a
detailed analysis \cite{boundstate} of this fine-tuning requirement and
obtained the estimated value $(g_t)_{MPP} = 1.00 \pm 0.14$, which within errors is
in agreement with the experimental value of the top quark mass. However this result is
controversial, as it has not been confirmed by other calculations
\cite{Kuchiev}.

With our assumption of the
Multiple Point Principle, it is of course much easier to get several
phases of the vacuum realized, since they can now coexist in balance
w.r.t. energy density. Indeed in our model we have two vacua, with
and without the $6t + 6\overline{t}$ bound state condensate. So, at the
temperature scale of the weak interactions in the early universe,
there would be a random distribution in space of these two vacuum phases.
In the present article we estimated that the phase without a condensate would
eventually dominate, but that balls of the phase
with a condensate could be prevented from contracting for some time.
This is due to an effect of the lower Higgs field inside the balls
(i.e. in the with condensate phase) causing lower masses for
several Standard Model particles,
which in turn cause them to provide a bigger pressure through their
Planck radiation.
It follows that many balls would remain essentially
uncontracted until the temperature fell below 10 MeV. At this
temperature a stabilisation process set in, due to the balls getting
filled with nucleons, which stopped the balls from totally contracting
away. In other words we actually estimated that, under the
conditions derived from just the Standard Model with the MPP determined
couplings, permanently stabilised balls would be formed with an
excess of nucleons inside. Really it is the vacuum inside the balls
having a bound state condensate which captures at first almost all  the nucleons.

It is speculated that an explosive fusion of
helium to heavier nuclei occurs inside these ``with condensate'' balls,
causing about 1/6 of the nucleons to escape outside into the dominant
``without condensate'' vacuum. We
estimated that these processes of contraction of balls and
their explosive emission of nucleons
would be mainly finished, at a temperature of about 2.3 MeV, just
in time for {\em not} disturbing the Big Bang Nucleosynthesis.
After that time the balls have
their sizes stabilised
by their content of ordinary matter. Then they are so small and
concentrated with density around $10^{14}$ kg/$\hbox{m}^3$ that by far
they dominantly only interact by gravity. Thus the balls
can be treated as just dark matter from there on. Of course if
you get really close to a
single ball, like in Tunguska, its non-gravitational interactions
can become relevant. In this way we
managed, for practical purposes, to obtain a dark matter model
from just the Standard Model extended with the extra assumption of
the Multiple Point Principle!

In addition this dark matter model
leads to events of the Tunguska
type, one about every 200 years, and each of them would dig deep holes into the earth.
It is not excluded that they would go out again at another place on the earth, since
the order of magnitude for the penetration depth of a
dark matter ball is similar to the earth radius.
These holes would of course be filled with molten material, which is
likely to be  pumped up from much deeper regions in the earth. In fact they
would be much like kimberlite pipes. We are therefore tempted
to identify these pipes, formed by the balls falling through the earth,
with kimberlite pipes. Indeed we made a model for estimating the number
of such kimberlite pipes which should be exposed so as to be visible
on, for example, the Canadian Shield. We found that, within the large
uncertainties of our estimate, the number of kimberlite pipes
actually found were compatible with our estimate of the number
expected from dark matter ball impacts with the earth.
Now, however, it must be
admitted that we have a problem for this picture of the origin
of kimberlite pipes.
The times of creation of various kimberlite pipes have been
estimated \cite{Jelsma,Heaman} and a distribution found, even in
regions where Archean pre-Cambrian rocks were exposed,
with clustering in various geological eras. In particular
there is a concentration of kimberlite emplacements in
a couple of clusters during the Cretaceous period earlier
than about 200 million years ago. This is in contradiction
with our picture, because in our model we must have equally
many Tunguska-like impacts per year
both during the very long pre-Cambrian eras and the comparably
shorter Cretacious and Mesozoic eras.
So, if Archean rocks are exposed, why should only the kimberlite pipes from a
much later time be visible there? This is strange in our model, in which
Tunguska-like particles fall at random and only
the visibility today can be dependent on the geological
conditions. There should be no connection in our model
with tectonic events like the splitting up of the supercontinent Gondwanaland.

If the kimberlites should match our model too badly, there is of course the
possibility that our falls of dark matter as Tunguska events have nothing
to do with kimberlites.  Potentially our ball falls could even
be associated with one special kind of kimberlite, but
there may be so few of them that they would be
hidden among the ``normal'' kimberlites.

Apart from the problem with the time distribution of the kimberlite pipes
we have a very consistent and viable picture of both dark matter
and the Tunguska event.
We present   the values of the various parameters in our model in Table 1.

\begin{center}
\begin{table}[h]
\caption{The parameters of our model picture of the Tunguska particle
as a ball of a new type of vacuum with a bound state condensate, filled
with ordinary white dwarf-like matter and on the borderline of stability.}

    \begin{tabular}{ | l | l | l | l |}
\hline
    Time Interval of impacts & $r_B^{-1}$ & 200 years &\\
\hline
    Rate of impacts& $r_B$& $1.5 *10^{-8}\ \hbox{s}^{-1}$  &  \\ \hline
    Dark matter density in halo & $\rho_{halo}$ &0.3 GeV/cm$^3$ &\\ \hline
    Dark matter near solar system & $\approx 2\rho_{halo}$ &0.6 GeV/cm$^3$ &  \\ \hline
    Mass of the ball & $m_B$ & $1.4*10^8$ kg &  \\
    \hline
Estimated typical speed of ball& $v$& 160 km/s& \\ \hline
Kinetic energy of ball & $T_v$ & $1.8*10^{18}$ J&  430 megaton TNT \\\hline
Energy observed in Tunguska& $E_{Tunguska}$ & $(4-13)*10^{16}$ J  &
10-30 megaton TNT\\ \hline
 Potential shift between vacua& $\Delta V$ & 10 MeV & \\ \hline
Cube root of tension & $S^{1/3}$& 28 GeV& from $m_B$ and rate\\ \hline
Cube root of tension& $S^{1/3}$ & 16 GeV & from condensate \\ \hline
Ball density& $\rho_B$ & $10^{14}$ kg/m$^3$ & \\ \hline
Radius of ball & $R$ & 0.67 cm& \\ \hline

\end{tabular}
\label{table}
\end{table}
\end{center}

All together we think that our model - apart from troubles with
the time distribution of kimberlite pipes, and some dispute
\cite{boundstate, Kuchiev}}
about the predicted top quark Yukawa coupling  - has many features that
makes it viable as a picture explaining at first sight rather
separate phenomena in high energy physics, the Tunguska event,
geology and cosmology.

The most favoured competing model for the explanation of the
Tunguska event is the comet hypothesis.
However, in order to explain the lack of iridium from the Tunguska
event, the comet would have to have been especially low in dust
content. Halley's comet has $~40\%$ dust and such a comet would
have given
too much iridium and
also presumably left more debris from the comet on the site. Really the
most remarkable fact about the Tunguska event is the lack of any
rudiments of the impact object itself. That is why our model, in
which the cosmic body went so deep into the earth
that it practically disappeared, has a
good chance to be true (of course ice may also disappear).

Now what could give a hint as to whether our model is indeed
true?:

\begin{itemize}
\item{New Bound State}

If the high energy physics part of our model is indeed true, there
should be at least a couple of phases of the vacuum, in one of
which there is a condensate of a bound state of $6t +
6\overline{t}$. This bound state is then, by our Multiple
Point Principle, tuned in to be between being a tachyon and an
ordinary positive mass squared particle. Hence its mass must have been
fine-tuned to be rather small compared to the typical mass of 12
top quarks, which is 2 TeV. We have already suggested \cite{nbs,nbslhc} that
this bound state could be produced in pairs in
co-production with top-quarks in the LHC-accelerator.
Indeed we have estimated the mass of this new bound state in
appendix \ref{mbs} to be about $m_{NBS} \approx 260$ GeV.
Hence a pair of of new bound states cannot contribute to the
invisible decay width of a Higgs particle of mass 126 GeV,
although they would couple strongly to it.

\item{Other Dark Matter Models should be false}

It is of course clear that a likely way to falsify our model
would be to establish another model for dark matter, such as
a weakly interacting massive particle (WIMP) often identified as
the lightest superpartner in supersymmetric models. The DAMA
collaboration \cite{Dama} have claimed a positive signal for
direct WIMP production from the observed annual modulation in
their data over 12 annual cycles. More recently the COGENT
\cite{Cogent} and CRESST \cite{Cresst} collaborations have
also claimed evidence for WIMP production. However it is very
difficult to reconcile these positive results with the negative
results from the CDMS, EDELWEISS, XENON and LUX experiments
\cite{pdg, Cdms}.

\item{Lack of New Physics at LHC}

Our model is based only on the Standard Model. Therefore the LHC
gives support to our model, as long as it continues to provide no
evidence for new physics with a suitable dark matter candidate.

\item{Absence of MACHOs}

Microlensing searches \cite{pdg, Machos} for dark matter in the form
of massive compact halo objects (MACHOs) are insensitive to objects
with masses less than $10^{-8}M_{\odot}$. Hence our dark matter balls
are too light for observation by microlensing. In fact observations
\cite{Machos} show that MACHOs contribute less than $8\%$ to the mass
of the galactic halo and hence do not provide a significant source of
dark matter.

\item{Looking for effects of our balls in stars}

It is possible that dark matter balls collect into the core of a
collapsing star. Then, when the density and temperature in the interior
of the star gets sufficiently big, the balls could catch the nucleons,
particularly any free neutrons since they do not feel the electric
potential which has to be passed to get into the vacuum with a condensate
inside the ball. This electric potential is of the order of 10 MeV
and lower for the balls bigger than the minimal size, which we have
taken to be close to the typical size. Thus the physics of supernovae
could be changed by our model, if some balls expand and finally
make the resulting neutron star into what is formally really a
huge piece of dark matter.

\end{itemize}

\section{Acknowledgements}

CDF would like to acknowledge the hospitality and support from
Glasgow University and the Niels Bohr Institute. HBN would like
to acknowledge the hospitality and his status as professor emeritus
at the Niels Bohr Institute.

\appendix
\section{Appendix: Properties of condensate}
\label{vacuumstructure}

In order to estimate especially the surface tension, but also the
attraction potential $\Delta V$ for nucleons into the vacuum with
the condensate, we shall now set up a crude picture of the
condensate of the bound states of 6 top + 6 anti-top quarks. The
12 quarks form a closed 1s shell in this ``new bound state''
(NBS). The crude assumption to initiate our estimates is to say
that, instead of only thinking of the condensate as consisting of
a region with the bound states present, we can also imagine that
one bound state would begin to attract further top and anti-top
quarks in addition to the first 12; now they should collect in the
2s and 2p states. Really our hypothesis is that the distances
between the bound states in the condensate can be estimated, by
assuming the nearest bound states to a given one to be at the
distance from the centre of the bound state out to a top quark in
a 2s or 2p orbit around the bound state. In appendix C of our
previous article \cite{boundstate} we obtained an estimate of the
radius $r_0$ of the bound state:
\begin{equation}
r_0 \approx \sqrt{3/4}\frac{1}{m_t},
\end{equation}
where $m_t$ is the top quark mass. Actually $r_0$ was defined by
assuming that the single particle wave function for a top or
anti-top quark in the 1s bound state takes the form $\psi \propto
\exp(-r/r_0)$; so the mean square radius is given as $<r^2>= 3
r_0^2$.
Remembering that the Bohr radius for an  orbit with principal
quantum number $n$ is proportional to $n^2$, we get that the next
i.e.~$n=2$ orbits have a radius of about $r_{n=2} = 4r_0$. So,
with our crude assumption, the distance between ``neighbouring''
bound states should be about $4r_0$. In order to have the correct
number of top quarks for each colour and spin in such a region at a
distance $4r_0$ away from a bound state, we should have four
``neighbouring'' bound states at this distance of $4r_0$. This
brings to mind the distribution of carbon atoms in a diamond, in
which each atom is surrounded by four nearest neighbours. Now
diamond has a bond-length of $1.54 *10^{-8}$ cm and has a density
of $1.76 *10^{23}$ $\hbox{cm}^{-3}$; so in a cube, with a side of
length equal to the bond-length, there are 0.64 carbon atoms.
Taking our condensate to also have this property, we estimate that
our condensate has a density of
0.64 bound states per cube with side of length equal to the
bond-distance, which we took to be $4r_0 \approx 4 \sqrt{3/4}
/m_t$. Hence the density of bound states in the vacuum with a
condensate is estimated to be
\begin{equation}
 \rho_{number} = 0.64 * \frac{1}{(4 r_0)^3}
=0.0154 m_t^3=
 (43\ \hbox{GeV})^3. \label{numberdensity}
\end{equation}

Let us introduce an effective scalar field $\phi_{NBS}$ for the bound state.
Now $\phi_{NBS}$ is a real field and one cannot define a conserved
particle number. So, purely for the purpose of introducing a
number density of bound states $\rho_{number}$, we will formally
treat $\phi_{NBS}$ as a complex field in the following definition
of the number of bound states in a given volume:
\begin{equation}
 N = \int \phi_{NBS}^{\dagger} \stackrel{\leftarrow \rightarrow}
 {\partial_0} \phi_{NBS} \, \mathrm{d}^3\vec{x}.
\end{equation}
Denoting the expectation value of $\phi_{NBS}$ in the condensate by
\begin{equation}
 v = <\phi_{NBS}>
\end{equation}
and the energy of a bound state in the condensate by $E_{cond}\approx m_{NBS}$,
our estimate of the number density then becomes
\begin{equation}
 \rho_{number} = 2E_{cond} v^2.\label{wrong}
\end{equation}
Here $m_{NBS}$ denotes the mass of our new bound state in the
vacuum with a condensate.

Our vacuum with the condensate is supposed to be
a relativistic invariant vacuum - rather than an ether-like state
representing a special frame. Thus, strictly speaking, we must
imagine that our diamond model is moving with a superposition of
velocities having a Lorentz invariant distribution. That might
give divergences, if considered seriously, because of the infinite
Haar-volume of the non-compact Lorentz group. But let us hope the
divergences will cancel out and consider the case when the speed
of the crystal of bound states (more realistically one should of
course treat the diamond-like structure as a fluid)
has an associated $\gamma$. Then the energy of the bound state
$E_{cond} = \gamma * m_{NBS}$, but also the whole ``crystal'' gets
Lorentz contracted and its thickness in the direction of motion is
diminished so as to be $1/\gamma$ times as thick as without the
motion. The number density we calculated in (\ref{numberdensity})
i.e.~$\rho_{number}=(43 \ \hbox{GeV})^3$ would, if boosted by this
Lorentz contraction, increase its value to $\gamma *\rho_{number}
= \gamma * (43 \ \hbox{GeV})^3$. Allowing for this boosting and
using (\ref{numberdensity}), the correct form for (\ref{wrong})
becomes rather
\begin{equation}
\gamma \rho_{number} = \gamma* (43 \ \hbox{GeV})^3 = \gamma 2 m_{NBS} v^2
\end{equation}
or equivalently
\begin{equation}
 \rho_{number} = (43 \ \hbox{GeV})^3 =  2 m_{NBS} v^2.\label{right}
\end{equation}

\subsection{Effective potential for the effective field $\phi_{NBS}$}
\label{potential}

In order to facilitate the estimation of the surface tension in
the surface between the two phases of the vacuum, we shall now
introduce an ansatz for the effective potential for the effective
bound state field $\phi_{NBS}$. The basic assumption underlying
our model is the existence of these two degenerate vacua in the
Standard Model. Hence the effective potential for $\phi_{NBS}$
must have two degenerate minima. We shall take the effective
potential to be a function of the squared field $\phi_{NBS}^2$.
The philosophy for only taking the squared field in the effective
potential is that it should be easier to produce a pair of bound
state out of the vacuum than to make a single bound state, so to
speak, from its complicated 12 quark substructure. Thus we expect
to get pairs of bound states, as if they had an approximately
conserved quantum number taking values in $Z_2$ (the integers
modulo 2). In any case we hope that the choice of ansatz for the
effective potential should not make so great a difference order of
magnitudewise. We shall use a polynomial ansatz and take the
lowest order polynomial compatible with having two degenerate
minima.  So we must take the sixth order polynomial (meaning
third order in $\phi_{NBS}^2$):
\begin{equation}
 V_{eff}(\phi_{NBS}) = \frac{1}{M^2}\phi_{NBS}^2(\phi_{NBS}^2 - v^2)^2.
\label{Veff}
\end{equation}
This potential has been arranged to have two degenerate minima,
namely for $\phi_{NBS} = 0$ and for $\phi_{NBS} = \pm v$ (in a way
even three minima if you consider $ -v$ and $v$ as different.) The
potential has a maximum for $\phi_{NBS}= \pm v/\sqrt{3}$. The idea,
of course, is that in the phase in which we live and there is no
condensate we have the VEV $<\phi_{NBS} > = 0$, while in the
phase inside the balls where there is a condensate of the bound
states we have say $<\phi_{NBS}> = v$. The two values of the mass
squared $m_{NBS}^2$ of the bound state, as would respectively be
observed by observers living in these two phases, are obtained as
the second derivatives of the effective potential
$V_{eff}(\phi_{NBS})$ at the two minima:
\begin{eqnarray}
 \left. \frac{\partial^2 V_{eff}}{\partial \phi_{NBS} ^2} \right|_{\phi_{NBS} = 0}
 & = & \frac{2v^4}{M^2} = m_{NBS}^2|_{without \ condensate}
 \label{mwithout}\\
 \left. \frac{\partial^2 V_{eff}}{\partial \phi_{NBS} ^2} \right|_{\phi_{NBS} = v}
 & = & \frac{8v^4}{M^2} =
m_{NBS}^2|_{with \ condensate}. \label{mwith}
\end{eqnarray}
Thus, with our ansatz (\ref{Veff}), we obtain
\begin{equation}
m_{NBS}|_{with \ condensate} = 2 m_{NBS}|_{without \ condensate}.
\label{meq}
\end{equation}

Substituting the value of the mass of the bound state in the
``with condensate phase'' (\ref{mwith}) for $m_{NBS}$ in formula
(\ref{right}) gives us
\begin{equation}
 \rho_{number} = (43\ \hbox{GeV})^3 = 2 m_{NBS}|_{with \ condensate} v^2 =
\frac{4\sqrt{2}v^4}{M}. \label{rel}
\end{equation}

\subsection{The solitonic wall and surface tension}
\label{solitonic}

At the borderline between the two phases the field must, over a
rather short distance, go from $\phi_{NBS} = 0$ to $ \phi_{NBS} =
v$ as a soliton. Our surface tension or surface energy per unit
area $S$ is given by the Hamiltonian density for a static field
configuration
\begin{equation}
 {\cal H}_{static} = \frac{1}{2}
 \left( \frac{\partial \phi_{NBS}}{\partial x^i}\right )^2  + V_{eff}(\phi_{NBS})
\label{staticH}
\end{equation}
evaluated for the static soliton solution. The equation for the
soliton solution considered in only one dimension $x$, say,
perpendicular to the wall can be written
\begin{equation}
 \frac{1}{2} \left ( \frac{\partial \phi_{NBS}}{\partial x} \right )^2
 - V_{eff}(\phi_{NBS}) = C.
\label{sol}\end{equation} The constant $C$ is easily seen to be
zero, by considering the field $\phi_{NBS}$ a long away inside the
phases.
So a crude typical value for the slope $\frac{\partial
\phi_{NBS}}{\partial x}$ inside the solitonic wall is determined by the
square root of the maximum of the effective potential between the
two degenerate minima. This maximum value occurs at the peak of the
potential, where $\phi_{NBS} = v/\sqrt{3}$,
\begin{equation}
 V_{eff}|_{peak} = V_{eff}(v/\sqrt{3}) = \frac{4}{27}\frac{v^6}{M^2}. \label{peak}
\end{equation}
At the peak of the effective potential the gradient of the soliton
solution $\frac{\partial \phi_{NBS}}{\partial x}$ is equal to
$\sqrt{2}$ times the square root of $V_{eff}|_{peak}$:
\begin{equation}
\left. \frac{\partial \phi_{NBS}}{\partial x} \right|_{peak} = \sqrt{\frac{8}{27}} * \frac{v^3}{M}.
\end{equation}
A crude estimate of the thickness $d$ of the solitonic wall is
given by multiplying the inverse of this gradient by the
difference between the values of $\phi_{NBS}$ in the two vacua,
which is just $v$. So we get the thickness to be
\begin{equation}
 d= v\sqrt{\frac{27}{8}} * \frac{M}{v^3} =
 \sqrt{\frac{27}{8}} * \frac{M}{v^2}\label{d}.
\end{equation}
Now the energy density inside the soliton is given by
(\ref{staticH}), in which the two terms are equal to each other.
This follows from (\ref{sol}) with the constant $C = 0$. Averaged
crudely over the soliton, we can take the soliton energy density
to be half the peak value of the Hamiltonian density
$\frac{1}{2}{\cal H}_{static}|_{peak} = V_{eff}|_{peak}$, which is
given by (\ref{peak}).
So the energy per unit area of the surface or wall becomes this density multiplied by
the thickness $d$,
\begin{equation}
 S = d * V_{eff}|_{peak} =
\frac{\sqrt{2} v^4}{3\sqrt{3}M} = \frac{1}{12\sqrt{3}} * \rho_{number}=
(16\ \hbox{GeV})^3,\label{long}
\end{equation}
where we have used (\ref{numberdensity}).

\subsection{Mass of bound state NBS}
\label{mbs}

In order to also extract the mass of the bound state, we need
one more relationship between the parameters in the effective
potential $V_{eff}$. We do this by estimating the thickness of the
solitonic wall, using the properties of the condensate. We again
suppose the condensate has a density of bound states given by
taking the distance between neighbouring bound states to be of the
order of the radius of the $n=2$ orbit. The thickness of the wall
must be of the order of the thickness of the atomic layers in the
assumed diamond-like crystal structure. We shall take this layer
thickness to be given by half the bond-length between two nearest
neighbours, which is supposed to be $r_{n=2} =4r_0$. Then we
obtain the following
``physical'' estimate  of the wall thickness $d$,
\begin{equation}
 d \approx
2r_0 =
\frac{\sqrt{3}}{m_t} = \frac{1}{100\ \hbox{GeV}}.
\end{equation}
Combining this thickness estimate with the formula (\ref{d}) we get
\begin{equation}
 \frac{1}{100\ \hbox{GeV}} \approx \sqrt{\frac{27}{8} } * \frac{M}{v^2}.
\end{equation}
Combining this with (\ref{long}),
\begin{equation}
 \frac{\sqrt{2} v^4}{3\sqrt{3} M} = (16\ \hbox{GeV})^3,
\end{equation}
we get the following values for the parameters in our effective potential:
\begin{eqnarray}
 M &= & 0.45\ \hbox{GeV}\\
v &=& 9.1\ \hbox{GeV}.
\end{eqnarray}
Then, using (\ref{mwithout}, \ref{mwith}), we obtain an estimate for the
mass of the bound state in our vacuum and in the vacuum with a condensate:
\begin{eqnarray}
m_{NBS}|_{without \ condensate}&\approx& 260\ \hbox{GeV}\\
m_{NBS}|_{with \ condensate} &\approx& 520\ \hbox{GeV}.
\end{eqnarray}

\subsection{Estimate of Higgs field in the condensate}
\label{vev2}

In section \ref{qmd} we estimated the value of the
parameter $\Delta V$ to be 10 MeV, by assuming that the value of
the Higgs field $<\phi_h>$ in the condensed phase is just 1/2 of
the usual Higgs field VEV in the no-condensate phase. However
there is a large uncertainty on the value of the ratio of the VEVs
of the Higgs field in the two vacua.
Here we want to make an estimate of this ratio of VEVs using the properties of
the condensed phase conceived of as having a diamond-like structure, although
really being a fluid.

First notice that we have a number density $\rho_{number} = (43\
\hbox{GeV)}^3$ for our new bound states (NBS) present in the
condensed phase. Around each of these bound states, we have the
negative Yukawa potential contribution to the Higgs field
\begin{equation}
 \Delta\phi_h|_{\hbox{one NBS}} = \frac{12 g_t/\sqrt{2}}{4 \pi r}* \exp(-mr),
\end{equation}
where $r$ is the distance from the centre of the NBS to the point
at which we want the contribution $\Delta \phi_h$ to the Higgs
field.  The effective or true Higgs mass is here denoted as $m$.
Averaging over space of course means
\begin{equation}
 av(...) = \frac{ \int ... \, \mathrm{d}^3\vec{x}}{V} =
 \frac{ \int ... \, \mathrm{d}^3\vec{x}}{\int \mathrm{d}^3\vec{x}},
\end{equation}
where $V$ is the infrared cut-off volume of the universe. Now we
want to sum over the contributions from all the $V\rho_{number}$
bound states in the infrared cutoff region $V$. Then we calculate
the average $\Delta \phi_h$ summed over all the bound states, so
as to obtain the total reduction of the ordinary Higgs field
expectation value $<\phi_h>_{no \ condensate \ phase}$ = 246 GeV
in the condensed phase. This average becomes
\begin{eqnarray}
 av(\Delta \phi_h|_{total})& =& \frac{1}{V}
 \int \sum_{bound \ states}\frac{6\sqrt{2}g_t}{4 \pi r}* \exp(-mr) \, \mathrm{d}^3\vec{x}\\
&=&
\frac{6\sqrt{2}g_t}{4\pi}* \frac{1}{V}\sum_{bound \ states}
\int_0^{\infty}\frac{1}{r}\exp(-mr) * 4 \pi r^2 \, \mathrm{d}r \\
&=& \rho_{number} 6\sqrt{2} g_t \int_0^{\infty} r \exp(-mr) \, \mathrm{d}r.
\label{suppression}
\end{eqnarray}
The Higgs field expectation value in the condensate phase then becomes
\begin{equation}
<\phi_h>_{condensate \ phase} = <\phi_h>_{no \ condensate \ phase} -
av(\Delta \phi_h|_{total}).
\end{equation}

Note that the expression (\ref{suppression}) is convergent
due to the Yukawa exponential factor $\exp(-mr)$.
This means that the contribution to the change in the Higgs field
from each single NBS bound state is only important over a region cut-off
by the Higgs mass exponential factor $\exp(-mr)$. Since the Higgs
mass effectively depends on the strength of the Higgs field,
we would prefer to interpret this factor $\exp(-mr)$ as symbolic for
a correction factor to the zero Higgs mass approximation, which we
now discuss more carefully. In our previous paper \cite{boundstate}
we estimated that, inside of a distance $1.58 r_0$ from the centre
of the bound state, the effective Higgs mass even becomes complex.
This is because the second derivative of the Higgs field effective potential
is negative, for the values of the Higgs field in that region, and thus
the mass squared of the effective Higgs vibrations are also negative there.
In such a region one gets a sine or a cosine factor rather than the
exponential factor $\exp(-mr)$ and we think it is best to take the
Higgs mass to be zero there. Thus the Yukawa potential solution to the
static Klein Gordon equation, for the Higgs field contribution due to the
NBS bound state under consideration,
first begins its usual exponential decrease at a
distance $r = 1.58r_0$ from the center. So we should rather  make the replacement
\begin{equation}
 \exp(-mr) \rightarrow \exp(-m(r-1.58 r_0) ) .\label{l106}
\end{equation}
for the exponential factor in the Yukawa potential. Even for $r
\ge 1.58r_0$ the mass parameter $m$ should rather be some
effective Higgs mass corresponding to the second derivative of the
Higgs potential for the value of the field $\phi_h$ present there.
Such an effective Higgs mass must at least be smaller than the
measured Higgs mass of 126 GeV. Really it interpolates between
values starting from being zero at $r=1.58 r_0$ and ending by
being at most 126 GeV \cite{lhc}.

We have already assumed an approximate diamond-like structure for
the condensate, with a separation distance of $4r_0$ between
neighbouring NBS bound states. So we are led to  a crude estimate
of the most important region in the integral (\ref{suppression})
as being around $r \approx 3 r_0$. Let us now identify our
modified exponential factor (\ref{l106}) at this typical value of
$r = 3r_0$ with the original Yukawa form, but with an effective
Higgs mass $m_{eff}$ inserted for $m$:
\begin{equation}
\exp(-m(r-1.58 r_0) ) = \exp(-m_{eff}r) \ \ \hbox{for}\  r=3r_0.
\end{equation}
The value of this effective Higgs mass is then
\begin{equation}
 m_{eff} \approx \frac{3 -1.58}{3} * 126\ \hbox{GeV}
= 60\ \hbox{GeV}.\label{effHiggs}
\end{equation}
So we should get a roughly correct estimate for the reduction in the
Higgs field $av(\Delta \phi_h|_{total})$, by inserting the value (\ref{effHiggs})
of 60 GeV for $m$ into (\ref{suppression}).
Inserting the mathematical evaluation
\begin{equation}
 \int_0^{\infty} r \exp(-m_{eff}r) \, \mathrm{d}r = \frac{1}{m_{eff}^2},
 \label{l108}
\end{equation}
into (\ref{suppression}) we get
\begin{eqnarray}
 av(\Delta \phi_h|_{total}) &=& \rho_{number} 6\sqrt{2}g_t *\frac{1}{m_{eff}^2}\\
&=& \frac{(43\ \hbox{GeV})^3 6\sqrt{2}g_t}{(60\ \hbox{GeV})^2}\\
&=& 174\ \hbox{GeV}.
\end{eqnarray}
Here we used $g_t = 0.93$ for the top quark running Yukawa coupling constant.
Hence we obtain the ratio of the Higgs field VEVs in the two vacua to be
\begin{equation}
 \frac{<\phi_h>_{condensate\ phase}}{<\phi_h>_{no\ condensate\ phase}} =
\frac{246 - 174}{246} \approx 0.3,
\label{vevratio}
\end{equation}
which only differs by a factor of 3/5 from the value 1/2
we used in section \ref{qmd} and throughout the paper.

We will now check the self-consistency of our assumption that, for a general point,
the typical distance to the nearest bound state is $3r_0$ and that
the integral (\ref{suppression}) is dominated by the region around $r \approx 3r_0$.
This dominant region should correspond to where the integrand of (\ref{l108}) has
its maximum value. This integrand $ r\exp(-m_{eff}r)$ has its maximal value for
$r = r_{max} = \frac{1}{m_{eff}} = 1/(60\ \hbox{GeV})$,
which agrees well with $3 r_0 = 3 * \sqrt{3/4}/m_t =1/(67\ \hbox{GeV})$.

\section{Appendix: Production of balls in early universe}
\label{appendix}
In this appendix we shall go through a few detailed points about the cosmological
development of our balls in the early universe.  First we shall discuss
whether we can arrange the  normal vacuum without the condensate to become dominant.
Secondly we discuss the distribution
of ball sizes.
As the third subject we present our estimates of the
effects of the lower Higgs field inside the balls, which provide a pressure
that pumps up the balls and allows them to survive long enough to collect nucleons.
Fourthly we must discuss the nuclear physics inside the balls and whether or not
the fusion of nucleons to helium, and then from helium to heavier nuclei,
can occur on the right time scales for our model to work.
In particular we consider whether the appropriate number of nucleons
could be expelled from the balls, in the explosive fusion of helium to
the heavier nuclei, to form what is now seen as ordinary matter.

\subsection{Which vacuum takes over?}
\label{whichvacuum}
At temperatures high compared to the weak
scale, meaning $T
\approx 100\ \hbox{GeV}$, there must be about equal volumes of the
vacuum {\em with} and the vacuum {\em without} the condensate. It
is at the time, when the temperature is of the order of the weak
scale, that walls between the regions with different vacua -
``with condensate'' or ``without condensate'' - begin to contract. This contraction
diminishes the local pieces of the vacuum with the larger free energy
density at the weak scale.
At higher temperatures the walls
were present by having effectively zero free energy density.  Now
the zero temperature energy density difference between the two
types of vacuum is, by our Multiple Point Principle assumption,
very small - essentially zero. Hence the free energy density
difference comes from those species of particles which behave with
different mass in the two different vacua.  For various
temperature scales most particles have a mass either bigger or
smaller by an order of magnitude or more than the temperature. If
that is the case for a certain species and the mass difference in
the two vacua is only of the order of the mass itself, then that
species is either effectively so heavy as to almost not be present
or it effectively has zero mass. In both cases the difference in
free energy density will be very small between the two phases from such a
species. So, in order to get an important difference in free energy density,
we must  call attention to just that or those species having their
mass of the order of the temperature. The most important choice
of species, for determining which of the two vacua shall contract,
is taken close to the beginning of contraction at the weak scale temperature era.
The particle that has the right mass to be important for the free
energy density difference in this era is the bound state (NBS) responsible for the
condensate itself. This is because the mass $m_{NBS}$ of the
bound state is connected to the same
effective potential that gives the wall its energy density, which
is of course really the scale that decides when the walls between
the vacua begin to contract.
We have considered an ansatz for this effective potential $V_{eff}(\phi_{NBS})$
in appendix \ref{potential}, where we found (\ref{meq}) that the mass in
the vacuum with the condensate $m_{NBS}|_{with\ condensate}$ is larger
than the mass in the normal vacuum without a condensate
$m_{NBS}|_{without\ condensate}$.
So, when the temperature is in between the two masses
$m_{NBS}|_{without\ condensate}$ and $m_{NBS}|_{with\ condensate}$,
there is
approximately Planck radiation of this bound state in the lower
mass vacuum - being the normal one without a condensate - but
virtually absent in the higher mass vacuum. Now such
Planck radiation gives a {\em negative} free energy density. Thus
finally the free energy is minimised by having as much as possible
of the normal phase without a condensate. This is supposed to
initiate the contraction in the direction of diminishing the
amount of vacuum with a condensate. That should be the reason why we today
have ended up with dominance of the without condensate phase and only find the with
condensate phase inside tiny balls making up the dark matter.

\subsection{Distribution of ball sizes}
\label{distribution}

In principle the initial distribution of ball sizes is given by the Boltzmann
distribution of the walls at the effective starting time of the weak scale temperature.
A priori the situation could even be that the two phases penetrate each other
as two connected regions (with only small pieces of isolated balls).
However, at least after one
of the phases has taken over even if only by of order unity in relative volume,
there will be disconnected balls of the minority phase.
To obtain the distribution of the sizes - volumes or radii as if the balls were
already spherical (which they are not at first) - of these minority phase balls
a computer simulation should in principle be applicable. In any case it must be so
that the total volume of the minority phase balls per volume of space must be less
than half the space volume.
Letting the
number of (minority phase) balls per unit volume
with radius between $R$ and $R + \mathrm{d}R$  be denoted by
$\rho_{balls}(R) \, \mathrm{d}R$
this requirement comes to mean
\begin{equation}
 \int_0^{1/H} \rho_{balls}(R)\frac{4\pi}{3} R^3 \, \mathrm{d}R  <1/2.
\end{equation}
Here the upper limit on the size of the balls \cite{Kibble} at the
weak scale is given by the Hubble distance at that time $1/H
\approx 1\ \hbox{cm}$. This integral should converge.
So $\rho_{balls}(R)$ must at least fall  off asymptotically as
$1/R^4$ for large $R$ and for small $R$ it cannot grow towards
$R=0$ faster than also $1/R^4$. Under for instance a simple Hubble
expansion, the total density of balls per volume must decrease as
$T^{3}$ but at the same time the radius of a given ball will
increase proportional to $1/T$. Under an expansion corresponding
to the radiation domination temperature ratio $T_1/T_2$, we therefore
have the following change in the just defined ball density $\rho_{balls}(r)$
\begin{equation}
 \rho_{balls}^{T_1}(R) \rightarrow  \rho_{balls}^{T_2}(R)
 = \rho_{balls}^{T_1}(R*\frac{T_2}{T_1})*
\frac{T_2^3}{T_1^3} *\frac{\mathrm{d}R_1}{\mathrm{d}R_2 }
= \rho_{balls}^{T_1}(R*\frac{T_2}{T_1}) *\frac{T_2^4}{T_1^4}.
\end{equation}
The most crude just barely converging power law behaviour $\rho_{balls}(R) \propto 1/R^4$,
which we shall use, would
not change with time, as is easily seen, under such a Hubble expansion. So, as the Hubble expansion
goes on, we do not expect the distribution of ball sizes to change much apart from the possible
collapses discussed in appendix \ref{survival}. The Kibble upper radius bound \cite{Kibble} given
by the Hubble distance of course expands.
As cosmological time progresses the smaller balls contract first and then
the bigger and bigger balls successively contract.
So the $\rho_{balls}(R) \propto 1/R^4$ distribution gets
cut off at the small $R$ side by a time dependent cut-off.

Those balls which are sufficiently big to get their collapse halted by the nucleons
being pressed together inside them finally end up\footnote{Here we ignore the nucleons
being pumped in from smaller totally collapsing balls.} having masses $M$ proportional
to the volume {\em before the collapse}.
Thus we have $M \propto R^3$, where $R$ is taken at the time
when the temperature is $T=10\ \hbox{MeV}$. This leads for
the $\rho_{balls}(R) \propto1/R^4$ distribution to a mass
distribution $\propto \mathrm{d}M/M^2$ and the total average mass becomes a logarithmically divergent
expression cut off by the Kibble upper bound. So indeed it is not so bad
an approximation to consider it that the balls making up the dark matter have
a typical well defined mass, as we did in our calculations.
We can in fact estimate the median size of a dark matter ball, using the power law
distribution $\rho_{balls}(R) \propto 1/R^4$ cut-off at small R by the stability
borderline radius $R_{border}$ and at large R by the Kibble radius $R_{Kibble} = 1/H$.
Then the median ball radius is determined by the equation:
\begin{equation}
  \int_{R_{border}}^{R_{median}} \rho_{balls}(R)\, \mathrm{d}R =
  \frac{1}{2} \int_{R_{border}}^{R_{Kibble}} \rho_{balls}(R)\, \mathrm{d}R
\end{equation}
which gives
\begin{equation}
 \frac{1}{R_{border}^3} - \frac{1}{R_{median}^3} =
 \frac{1}{R_{border}^3} - \frac{1}{R_{Kibble}^3}.
\end{equation}
Since $R_{Kibble} \gg R_{border}$, it is a good approximation to take the limit
$1/R_{Kibble} \rightarrow 0$ giving
\begin{equation}
 R_{median} = 2^{1/3} R_{border}.\label{median}
\end{equation}
That is to say the typical or median ball radius $R_{median}$ is just a factor of
$2^{1/3}$ larger than the radius $R_{border}$ of the minimal stable ball. In this formula (\ref{median}) the radii involved were
really the radii {\em before the balls
collapsed}, so denoting that by an
assigned bracket would make this formula
read
\begin{equation}
R_{median}^{(before)} =2^{1/3}
R_{border}^{(before)}.
\end{equation}

Before the contraction we assume the
density of baryon number and thereby at
the end of mass per unit volume inside
the balls to be the same independent of
the size of the ball, so that the mass of
a ball
$m_B\propto
R^{(before)3}$. From (\ref{s2oR}), (\ref{ne}), (\ref{P}) and
(\ref{nne}) we get the number density of nucleons
 inside the ball after the contraction
\begin{equation}
n = \frac{2}{3\pi^2}\left( \frac{24\pi^2 S}{R^{(after)}}\right )^{3/4}.
\end{equation}
So
\begin{equation}
m_B \propto nR^{(after) 3} \propto R^{(after)
9/4}.
\end{equation}
But we also have of course
\begin{equation}
m_B\propto R^{(before) 3}.
\end{equation}
Hence
\begin{equation}
R^{(after) 9/4} \propto R^{(before) 3},
\end{equation}
and thus (\ref{median}) leads to
\begin{equation}
\frac{R_{median}^{(after)}}{R_{border}^{(after)}} = \left( \frac{R_{median}^{(before)}}{R_{border}^{(before)}} \right )^{4/3} =
(2^{1/3})^{4/3} =2^{4/9}.
\end{equation}
I.e.
\begin{equation}
R_{median}^{(after)}= 2^{4/9}R_{border}^{(after)}.
\end{equation}

When we ask for what fraction of the mass of the balls is in a given range of sizes
for a mass distribution $\propto \mathrm{d}M/M^2$,
we shall find that in each order of magnitude range you have equally much mass. But if
you ask for the number of balls, the majority will have masses close to the lower bound
given by the borderline for stability against collapse.
Thus a ball responsible for a random Tunguska-like event
would statistically have its mass not far from the stability borderline.

\subsection{Survival of Balls}
\label{survival}

After the era of the NBS which suppressed
the size of balls of the {\em with} condensate vacuum, as discussed in appendix \ref{whichvacuum},
a quite analogous effect sets in based
successively on the various quark,
lepton, Higgs  and gauge particle species.
These particles get their masses from the Higgs field
and thus somewhat different masses in the two
phases (vacua). For all the quark, lepton species and massive gauge particles and even the
Higgs\footnote{While, for the quarks, leptons and gauge particles, it is well-known
that their masses are proportional to the Higgs field expectation values,
in the case of the Higgs itself a moment more of contemplation is needed to see this. But indeed,
in going from one phase to the other, the Higgs self coupling coefficient $\lambda$
will stay practically constant and thus even for the Higgs we have such a formula.},
it is however so that the masses are smallest in the phase with
NBS condensation, because the Higgs
field expectation value
is the smallest in that vacuum.
(It is at least expected that in the condensate
phase with its high density of NBS states, inside which are highly reduced Higgs
fields, there will be a smaller Higgs field VEV. See appendix \ref{vev2}.) Thus the
effects from these quarks and leptons etc.,
oppositely to the effect from the NBS itself, go in the direction of making
the numerically largest but negative free energy density contribution in the
{\em with} condensate phase. It follows that the
effect of these Standard Model particles is to provide a pressure
expanding balls made from the {\em with}
condensate phase.
As temperature falls with time one
flavour of quark or lepton or gauge
particle  after the other
first gets less and less relativistic and at the end essentially disappears totally.
But, for each flavour or massive gauge particle, there is an interval of temperatures around the mass of
that particle  in which there are significantly more particles of this type
in the {\em with} condensate phase (inside our balls) than outside. This is because
the mass is smaller in this inside phase due to the smaller Higgs field expectation value there.
So when the temperature is in one of these intervals
there occurs, due to the particle in question, a more negative free energy
contribution in the inside or {\em with} condensate phase.
Then the minimisation of the free energy tends to
expand the balls due to this effect.
We assume there are sufficiently many Standard Model species
distributed smoothly enough in mass, so that the intervals of activity providing the pressure to
expand the balls this way worked all along the temperature scale from the weak scale to 10 MeV.
Clearly the expansion pressure is of the order of $T^4$, when in the
main region of the mentioned time intervals. Roughly we could imagine that the species
in question were already essentially extinct in the {\em without} condensate phase. It would only
be present inside the ball with an approximately massless Planck radiation pressure resulting.

Let us study the force per unit area on the wall between the two phases divided by
$T^4$ and give it the name
\begin{equation}
\sum_m F(T,m)= \frac{\hbox{``Planck pressure''}_{with \ condensate} -
\hbox{``Planck Pressure''}_{without \ condensate}}{T^4}.
\label{sumFTM}
\end{equation}
This is the contribution from all the degrees of freedom. The expressions for the ``Planck pressure"
from massless degrees of freedom are given in equations (\ref{pressureboson}) and
(\ref{pressurefermion}) of appendix \ref{neff}.
The symbol $\sum_m$ is supposed to mean the sum over the degrees
of freedom symbolized by the mass of the represented particles.
The division by $T^4$ is
of course really to extract a number, which essentially counts the number of degrees of freedom
contributing to the difference in force on the two sides of the wall.

The range of Standard Model masses extends over a logarithmic factor of
$\ln m_t/m_e \simeq 12.7$. These particles have 10 bosonic and 84 fermionic component
degrees of freedom. So there are $94/12.7 = 7.4$ components per $e$-factor in the mass range.
The pre-factor of $7/8$ for fermions and $1$ for bosons in the expressions
(\ref{pressureboson}) and (\ref{pressurefermion}) for the ``Planck pressure''
averaged over the 94 components is 0.89. We can now estimate the value of the expression
(\ref{sumFTM}), by replacing the sum by an integral over $\ln m$:
\begin{equation}
 \sum_m F(T,m) \approx 7.4\int F(T,m) \, \mathrm{d} \ln m
 \approx 0.89 *\frac{2}{3}* 7.4 \sigma \ln\frac{v_{without}}{v_{with}}.
\label{dFdlnT}
\end{equation}
Here $\sigma = \pi^2/60$ is the Stefan-Boltzmann constant, while
$v_{with}$ denotes the Higgs field VEV in the {\em with} condensate phase,
and $v_{without}$ is the Higgs-VEV in the {\em without} condensate  phase. The ratio of these
two expectation values of course leads to the same ratio for the masses between the two phases
for all the particles obtaining their masses from the Higgs field.

We now explain the origin of the factor $\ln\frac{v_{without}}{v_{with}}$ in equation
(\ref{dFdlnT}). Once we divide the $T^4$ factor out of the pressures,
the pressure on one side of the wall divided by the $T^4$ only
depends on the ratio of $T$ to the mass in the phase in question. The pressure from one species
on one side of the wall has, after the $T^4$ division out, the shape of an almost flat curve in the
high temperature region compared to the mass and is essentially zero on the low temperature side.
In order to construct (\ref{dFdlnT}) we need the difference between two graphs of this type.
For the average of this difference the exact form of the transition between the high temperature
and low temperature regions does not matter, since
it is the same for both the
mass in the with condensate phase and in the without condensate phase.
The logarithmic range between the low temperature cut-offs for a certain species,
in the without condensate phase and the with condensate phase respectively,
is of course $\ln\frac{v_{without}}{v_{with}}$.

Assuming a Higgs field ratio of $v_{without}/v_{with} = 2$ we get
\begin{equation}
 7.4\int F(T,m) \, \mathrm{d} \ln m \approx 0.89 * 2/3 * 7.4 *(\pi^2/60) \ln 2 = 0.50.
\end{equation}
Approximating the form of $F(T,m)$ by a delta function in $\ln(m/T)$ and,
assuming the function $T^4$ is smooth for this purpose,
we get the total force per unit area at the
temperature $T$ to be
\begin{equation}
 \sum_m T^4F(T,m)= 7.4\int T^4 F(T,m) \, \mathrm{d}\ln m = 0.50T^4.
 \label{sumT4F}
\end{equation}

In order that this pressure (\ref{sumT4F}) from the Standard Model particles should prevent a ball from
collapsing, it must be greater than the pressure $2S/R$ due to the surface tension.
Hence the radius $R$ of a ball, surviving (before being stabilised by collecting nucleons) to a temperature
$T$ without collapse, must be greater than a critical radius $R_{crit}(T)$
given by
\begin{equation}
 \frac{2S}{R_{crit}(T)} = 0.50 T^4,
\end{equation}
leading to
\begin{eqnarray}
 R_{crit}(T)
& =& \frac{4.0}{T}\left( \frac{28 \ \hbox{GeV}}{T}\right )^3.
\label{Rcrit}
\end{eqnarray}
For example at $T=10 \ \hbox{MeV}$ you obtain
\begin{equation}
 R_{crit}(10\ \hbox{MeV}) \approx \frac{4.0}{10\ \hbox{MeV}} 2800^3
\approx 2 \ \hbox{mm}.
\end{equation}
This means that even at the temperature scale $T=10 \ \hbox{MeV}$, when we can get balls supported from
collapse by their baryon content, the balls that can contract at all
are smaller than our ``observed'' ball, whose size at this stage is of order 6 m. Thus a
ball of the ``observed'' size really expands at this temperature rather than contracting!

\subsection{``Nucleon  conduction''}
\label{heatconduction}

 Our main suggestion is that at present the balls with an NBS-condensate are spanned out
by an excess of baryon number density inside the balls. Thus  it is
crucial for the functioning of our picture that in a temperature
interval around $T\approx 10$ MeV, when the Boltzmann distribution
can be hoped to arrange to get the baryon number concentrated in the
inside of the balls, such a concentration really happens.
Such a concentration can be imagined to occur at least in two ways:
\begin{itemize}
\item{1.} There is enough ``conductivity'' for this baryon number, so that
the baryon number distributes itself over the full interior
of the ball.
\item{2.} The walls run around so much in the $\Delta V = 10 $ MeV era that
they sweep up all the nucleons and get them into the
with condensate phase.
\end{itemize}
 For the purpose of possibility 1,  we must estimate the ``baryon conductivity'' $\kappa_b$
for the spreading of baryon number density in the (Planck) plasma.

We define a ``baryon number conductivity'' $\kappa_b$
so that the flow of the baryon number through the plasma $\vec{j}_b(\vec{x})$ is
given as
\begin{equation}
 \vec{j}_b = \kappa_b \nabla \rho_b(\vec{x}).
\end{equation}
Here $\rho_b(\vec{x}) $ denotes the number of baryons per cubic metre
minus the number of anti-baryons in the same volume around the point $\vec{x}$
(really we shall use it mainly here in the situation where there are
just nucleons and no anti-baryons). The constant coefficient $\kappa_b$
can be estimated in terms of the mean free path $\lambda_b$ and the
typical velocity $v_b$ for a nucleon in the plasma.

A nucleon hitting  a typical plasma particle, taken here to be
an electron or a positron, only gets scattered by a small angle.
The momentum transfer in such a small angle collision of the nucleon (proton) will
only be of the order of the momentum of an electron or positron
in the plasma at the temperature $T$. However a scattering by an
angle of order unity of say the proton requires a momentum transfer
of the order of $\sqrt{m_N*T}$. Using the hypothesis of the small angles
adding up like a random walk, we require that the momentum transferred
in the single collisions of order $T$ with random signs add up to
$\sqrt{m_N*T}$. That is to say we need of the order of
\begin{equation}
 n_{coll} \approx \left (\sqrt{\frac{m_N}{T}}\right )^2=m_N/T
\label{ncoll}
\end{equation}
small angle scatterings. We take the nucleon transport cross
section \cite{lifshitz} to be
\begin{equation}
 \sigma_b \approx \frac{g^4*2\pi\ln(1/\theta_{fcut})}{(4\pi)^2T^2 }\approx \frac{g^4}{8\pi}T^{-2}
\end{equation}
Here $\ln(1/\theta_{fcut})$ is the Coulomb logarithm\footnote{We shall ignore the Coulomb
logarithm here. Its effect would be to decrease our estimate of the baryon
number conductivity $\kappa_b$ and would, thereby, strengthen
our conclusion that the diffusion of nucleons in the plasma is too slow for them to
collect inside the balls.}, in which $\theta_{fcut}$
is the order of magnitude of the forward scattering angle below which, due to
Debye screening, the scattering can no longer be regarded as Coulomb scattering.
For the density of particles in the plasma we take
\begin{equation}
 n_{plasma} \approx n_s \frac34(\zeta(3)/\pi^2) T^3\approx 0.1 n_s T^3.
\end{equation}
So we end up with the following estimate of the mean free path for a nucleon
in the plasma:
\begin{equation}
 \lambda_b \approx \frac{n_{coll}}{\sigma_b n_{plasma}} \approx  \frac{n_{coll}8\pi}{n_s g^4T^{-2}*0.1*T^3}\approx
\frac{250 m_N}{n_sg^4 T^2}
\end{equation}
The typical velocity of a nucleon giving rise to the baryon ``conduction'' is
$v_b\approx \sqrt{T/m_N}$ and we note that it moves about a distance $\lambda_b$
in a random direction between collisions. So we estimate
\begin{equation}
 \kappa_b \approx v_b\lambda_b/3 \approx  \frac{80 \sqrt{m_N}}{n_sg^4 T^{3/2}}.
\end{equation}

On dimensional grounds, the time scale $t_b$ for getting the baryon number
smoothed out over a region, say a ball, of size
$R$ will then be of order
\begin{equation}
 t_b \approx \frac{R^2}{\kappa_b} \approx \frac{n_sg^4 T^{3/2}R^2}{80 \sqrt{m_N}}.
\end{equation}
If we here insert the relation $R\approx l_T= (6\ \hbox{m} *10 \ \hbox{MeV})/T= 3*10^{14}/T$
for the ball size in the era of temperature $T$ (before the ball collapsed)
from equation (\ref{lT}) valid for the typical
ball being present today in dark matter, we get
\begin{equation}
 t_b \approx \frac{n_sg^4(3*10^{14})^2 T^{-1/2}}{80 \sqrt{m_N}} \approx
\frac{n_s g^4* 7.5 *10^5}{\sqrt{m_N T}} \ \hbox{s MeV}.
\end{equation}
 For a temperature $T=10 \ \hbox{MeV}$ and say $n_s \approx 4$ this diffusion time for the baryon number becomes
\begin{equation}
 t_b \approx  \left \{ \begin{tabular}{cc c } 30000 \ \hbox{s}& for & $g^2=1$\\3\ \hbox{s} & for & $g^2=1/100$.
                            \end{tabular}\right .
\end{equation}
Comparing these numbers with the Hubble time at $T=10 \ \hbox{MeV}$ being 0.01 s, we see
that there is not sufficient time to have the baryon number just distribute over the
uncontracted balls by diffusion. So we seriously need the alternative 2 of wall motion,
which we consider in appendix \ref{fadingout},
in order to ensure the collection of the nucleons into the with condensate phase.

\subsection{Fading out of wall motion}
\label{fadingout}

Let us estimate the decay rate for a wave of definite wave number $\vec{k}$ of vibration
running along an otherwise imagined flat wall between the two phases of vacuum.
We are interested in how fast the inertia of the motion of the wall associated
with such a vibration gets damped out. Now, in the neighborhood
of the wall, there will appear a motion of the surrounding plasma partly following the motion of
the wall itself. In the approximation valid for our case
the equations describing the effective two dimensional motion of the plasma caused by viscosity,
due to the wave on the wall, are of the form that leads to harmonic functions.
At least one may easily estimate that the major layer of  moving plasma must
extend away from the wall a distance of the order of magnitude of the wavelength
of the wave on the wall.
Suppose, for example, that the plasma is only disturbed in a
thin layer (compared to the wavelength) along the wall. Then,
thinking of the plasma as an incompressible fluid, this plasma
material has to move much faster parallel to the wall than
perpendicular to the wall (in which direction it is enforced to move
to follow the wall). But this is intuitively unrealistic.
Also, if the extension of the disturbed plasma away from the wall were much longer than
the wavelength, it would mean that much bigger amounts of plasma would have to move than
needed if the penetration depth were only of the order of the wavelength.
Thus it follows that
the motion of the plasma
extends away from the wall by a distance of the order of the wavelength.
We assume - but it follows from very reasonable estimates about the
viscosity of the primordial plasma  - that the thickness of the plasma layer
crudely following the wall vibration is of the order of the wavelength of the wave
on the wall.

Letting $\mu$ denote the viscosity for the plasma,
the flow of momentum through the plasma is given by
\begin{equation}
 \vec{\vec{T}}(\vec{x}) = \mu \nabla \vec{v}(\vec{x}),
\end{equation}
where $\vec{v}(\vec{x})$ is the velocity field and thus a function of the position $\vec{x}$.
If the local velocity of the wall is $v_w$, the nearby plasma will of course follow it
by having $\vec{v}(\vec{x}) =  \vec{n} v_w$ near this point,
where $\vec{n}$ is a unit vector normal to the wall. This co-motion of the plasma
will die out over a distance of the order of the wavelength $\approx 1/k$ and thus the
taking of the gradient $\nabla$ essentially means multiplication by a factor $k$.
So, as time passes, the wall has to
deliver to the plasma a flow of momentum - that of course will damp its own
momentum density - per unit area of the order
\begin{equation}
 \hbox{``rate of loss  of momentum per unit area''}\approx -\mu |k| *v_w.
\end{equation}
 Since the mass density on the wall is $S$, this means an acceleration of the
wall (locally) given by
\begin{equation}
 S\dot{v_w} \approx - \mu |k| v_w.
\end{equation}
It follows that the motion of the wall $v_w$ relative to the bulk (far away)
plasma is being damped out with a decay rate $-\dot{v_w}/v_w$. Thus the
survival time for such an excess velocity $\tau_i$ or the survival time for
inertia of the wall is of the order
\begin{equation}
\tau_i = -v_w/\dot{v_w} \approx S/(\mu |k|)\approx
\frac{\hbox{``wavelength''} * S}{\mu }
\label{taui}
\end{equation}

Now the viscosity $\mu$ of the plasma is given as
\begin{equation}
 \mu \approx \rho * \frac{1}{2} \bar{u} \lambda
\end{equation}
where $\rho = n_s *\frac{\pi^2}{30}*T^4$ is the energy density,
$\bar{u}\approx  1$ is the root mean square speed of the
particles in the plasma and $\lambda$ is the mean free path in the plasma.
The density of particles in the plasma \cite{Kolb} is
$n = n_s*\frac{ \zeta(3)}{\pi^2}T^3$ where $n_s$ is the number of active species weighted with their
number of polarisations and a factor $3/4$ for fermions.  The cross section for
such a particle of the effectively
(relative to $T$) massless type and interacting
with a coupling constant $g$ behaves for dimensional reasons as
$\sigma \approx g^4 T^{-2}$. Thus the mean free path
in the plasma becomes
\begin{equation}
\lambda \approx  1/(n\sigma)
\approx \frac{\pi^2}{n_s\zeta(3)g^4T}
\end{equation}
and we obtain the following expression for the viscosity of the plasma
\begin{equation}
\mu \approx n_s\frac{\pi^2}{60}T^4*\frac{\pi^2}{n_s\zeta(3)g^4T} = \frac{\pi^4 T^3}{60\zeta(3)g^4}.
\end{equation}
The survival time (\ref{taui}) of the inertial motion
for a wave number $k$ thus becomes
\begin{equation}
 \tau_i
 \approx \frac{S*60\zeta(3) g^4}{\pi^4 T^3 |k|}
 \approx \left ( \frac{S^{1/3}}{T}\right )^3 * \frac{g^4}{|k|}
 \approx \hbox{``wavelength''} * g^4\left ( \frac{S^{1/3}}{T} \right )^3.
 \label{taui2}
\end{equation}

As an example let us consider waves on a typical ball (\ref{lT}) for which
the relevant ``wavelength'' is
\begin{equation}
 R\approx l_T= (6\ \hbox{m} *10\ \hbox{MeV})/T
\end{equation}
and estimate whether the waves would survive the Hubble time $t$.
The decay time due to
being stopped by the plasma for the waves obeying (\ref{lT}) relative to the Hubble time
is given by
\begin{equation}
 \frac{\tau_i}{t} = \frac{l_T*g^4S/T^3}{\hbox{MeV}^2\hbox{s}/T^2}=
g^4\left(\frac{2 \ \hbox{GeV}}{T}\right)^2.
\label{taut}
\end{equation}
We see that if $g^2\approx 1$, i.e. for a large $g^2$, the stopping power of the plasma is
sufficient to brake the wall movements in a Hubble time for temperatures above 2 GeV.
But, if the coupling squared is rather $g^2 \approx 1/100$, then the braking power will
be sufficient to stop the wall in a Hubble time for temperatures above 20 MeV.

It follows that the motion of the walls relative to the plasma will be stopped at the weak
scale. However they will be stopped in a curled up state and start moving again at a
lower temperature of $T = 2$ GeV for $g^2 = 1$ or $T = 20$ MeV for $g^2 = 1/100$.
So at these later times, especially in the 10 MeV era relevant for catching the nucleons,
we have rather freely moving walls.

This situation of rather freely moving walls is very important for the capturing of the
nucleons into the balls. The point is that, because of the clashing of the walls
against each other and crossing each other or being reflected against each other, almost any region of space
will get passed over  by some wall. Thus the nucleons anywhere
will meet a wall and get trapped on the side of the
wall with the condensate phase.
Thereby alternative 2 of appendix \ref{heatconduction} is justified to work, and all
the nucleons indeed get collected
into the with condensate phase (before the collapse of the balls).

\subsection{Effective number of degrees of freedom $n_{eff}$}
\label{neff}

In this subsection we shall estimate the value of the
critical size $R_{crit}(T)$ (\ref{Rcrit}) more accurately.
Especially we have to find out the value of the coefficient
$n_{eff}$ representing the number of active field components,
which can interact with the borderline wall and thereby stop the
contraction of the ball. For that calculation we must have in mind
that the wall between the two phases only interacts via the
variation in the Higgs field across the borderline surface.
Remember that throughout this paper we have assumed that
there is half as big a Higgs VEV in the
condensate phase as in the ``without'' condensate phase
(although our crude estimate (\ref{vevratio}) in
appendix \ref{vev2} gave a ratio of 0.3). But to
the majority of particles - the ordinary quarks and leptons - this
Higgs field only couples via the mass. So the interaction with the
wall comes only via the mass squared being different in the two
phases. When a particle tends to cross the wall it can quantum
mechanically risk to be reflected rather than, as happens under
the typical conditions of ultra-relativistic particles just
passing through the wall, being deflected only by a small angle.
As already explained, the parameter on which the reflection rate
for a particle $f$ depends is the change in mass squared of the
particle
\begin{eqnarray}
 \Delta m^2|_{\hbox{at wall}} &=&  g_f^2(<\phi_h>^2_{no \ condensate \ phase} -
<\phi_h>^2_{condensate \ phase}) \\
&= & g_f^2<\phi_h>^2_{no \ condensate \ phase} (1 - (1/2)^2) = \frac{3}{4}m_f^2.
\end{eqnarray}

Consider now a particle of type $f$ in an energy and momentum
eigenstate scattering on a wall, in say the xy-plane perpendicular
to the z-direction. Then the wave function for the situation in
which this scattering goes on continuously is of the form
\begin{eqnarray}
 \psi_{in}(\vec{r}) &=& \exp{i\vec{k}\cdot \vec{r}} + R * \exp{i(k_xx + k_yy -k_zz)}
\\
\psi_{out}(\vec{r}) &=& T * \exp{i\vec{k'}\cdot \vec{r}},
\end{eqnarray}
on the incoming and outgoing sides of the wall respectively. Here
of course $k_x' = k_x$ and $k'_y = k_y$. The coefficients $R$ for
the reflected wave and $T$ for the transmitted wave are to be
determined from the continuity conditions
\begin{eqnarray}
 1+R &=& T\\
ik_z -ik_zR &=& ik'_z T,
\end{eqnarray}
and the condition of the energy eigenstate being the same on both sides of the wall
\begin{equation}
 E = \sqrt{m_f^2 + \vec{k}^2} = \sqrt{m'^2_f +\vec{k'}^2}.
\end{equation}
This of course implies
\begin{equation}
 m^2_f + \vec{k}^2 = m_f^{'2} +\vec{k'}^2,
\end{equation}
and thus
\begin{equation}
 \Delta m^2 = (k_z - k_z')(k_z+k_z').
\end{equation}
Here of course say $m_f= g_f<\phi_h>_{condensate \ phase}$
and $m_f' = g_f <\phi_h>_{no \ condensate \ phase}$ or
oppositely. We easily find that the continuity equations transform to give
\begin{equation}
k_z-k_z' = (k_z+k_z')R,
\end{equation}
from where we get
\begin{equation}
 R= \frac{\Delta m^2}{(k_z + k_z')^2}.
\end{equation}
Thus the reflection  probability in such a scattering is given as
\begin{equation}
 |R|^2 =  \frac{(\Delta m^2)^2}{(k_z+k_z')^4 }.
\end{equation}

We see here that the rate of reflection cuts off rather quickly -
by a fourth power  - as the $k_z$ or $k_z'$ momentum becomes
bigger than  $``\Delta m'' = \sqrt{\Delta m^2}$. On the other
hand, at least if the $k_z$ for a particle on the way towards
scattering on the wall is so small that it is kinematically
impossible to enter the other phase, the reflection rate must of
course be $|R|^2 =1$. So when $k_z$ or $k'_z$ is of the order of
$``\Delta m''$ we expect to get reflection rates of order unity.
For large values of $k_z \approx k'_z$, our above estimate for the
reflection probability becomes
\begin{equation}
 |R|^2 \approx \frac{(\Delta m^2)^2}{16 |k_z|^4},
\label{reflection}
\end{equation}
an approximation that of course cannot be true unless $
\frac{``\Delta m''}{2|k_z|} < 1$. If this inequality is not
fulfilled, we may rather take it that
the reflection probability $|R|^2$ is of order unity.

It follows that the pressure exerted by a gas of such particles $f$ on the wall will
appear as if
only the fraction of the particles in this gas with
\begin{equation}
 |k_z| < \frac{``\Delta m''}{2}\label{gap}
\end{equation}
were present. We namely ignore the very rapidly falling off tail
of the distribution with a reflection probability
(\ref{reflection}) going as the inverse fourth power of $|k_z|$.
In fact we may crudely hope this may compensate for another of our
approximations being erroneous to the opposite side: we take the
reflection probability to be unity when $k_z$ obeys (\ref{gap}),
although of course there is a chance for the particle not to be
reflected.

Suppose now that the temperature $T$ is large
compared to the ``mass difference'' $``\Delta m''$.
Then, {\em if all the particles were unable to penetrate through
the wall but got reflected}, the gas would act like Planck
radiation and each polarisation component would exert a pressure
equal to 1/3 of the Planck energy density for one state of
polarisation. This ``Planck pressure'' is given as follows for
bosons and fermions respectively:
\begin{eqnarray}
`` Planck \ pressure''_b &=& \frac{2}{3} \sigma T^4 \ \
\hbox{for one-component bosons} \label{pressureboson}\\
``Planck \ pressure''_f &=& \frac{2}{3} *\frac{7}{8} \sigma T^4 \ \
\hbox{for one-component fermions},
\label{pressurefermion}
\end{eqnarray}
where $\sigma$ is the Stefan-Boltzmann constant
\begin{equation}
 \sigma= \frac{2k^4\pi^5}{15 c^2 h^3} = \frac{\pi^2}{60}.
\end{equation}
But now only  the particles with $p_z$ in a narrow interval around
zero of width $``\Delta m''$ contribute effectively to this
pressure. So, in the fermion case, the true effective pressure for
each component is reduced by the factor
\begin{eqnarray}
 \frac{ pressure_f}{``Planck \ pressure''_f}& =&
<``\Delta m'' \delta(k_z)>_{f; \ energy \ density \ weighted} \\
&  = &
\frac{ ``\Delta m''\int \delta(k_z) (\exp(\frac{|\vec{k}|}{T}) +1)^{-1}
*|\vec{k}| * \frac{\mathrm{d}^3\vec{k}}{(2\pi)^3}}
{\int (\exp(\frac{|\vec{k}|}{T}) +1)^{-1} *|\vec{k}|* \frac{\mathrm{d}^3\vec{k}}{(2\pi)^3}}\\
&=& \frac{``\Delta m'' \int_0^{\infty} (\exp(k/T )+1)^{-1}
2\pi k^2 \, \mathrm{d}k}{\int_0^{\infty}(\exp(k/T )+1)^{-1} 4\pi k^3 \, \mathrm{d}k}\\
&=& \frac{``\Delta m''*3/4*\zeta(3)\Gamma(3) }{2T*7/8*\zeta(4)\Gamma(4)}\\
&=& \frac{``\Delta m''90\zeta(3)}{T7\pi^4 } = \frac{``\Delta m''}{T} *0.159.
\end{eqnarray}
The corresponding reduction factor for bosons is given by
\begin{equation}
\frac{ pressure_b}{``Planck \ pressure''_b}  =
\frac{``\Delta m''\zeta(3) \Gamma(3) }{2T\zeta(4) \Gamma(4)}
= \frac{``\Delta m''}{T} *0.185.
\end{equation}

We can now define the effective number of degrees of freedom at
a temperature $T$ inside the ball for fermions
\begin{equation}
(n_{eff})_f = \sum_{relevant \ particles} g_f \frac{``\Delta m''}{T} *0.159 \label{f146}
\end{equation}
and for bosons
\begin{equation}
(n_{eff})_b = \sum_{relevant \ particles} g_b \frac{``\Delta m''}{T} *0.185.
\label{dofb}
\end{equation}
Here the sum over relevant particles means that only those particles, which have
sufficiently low mass as to be present in appreciable amounts at the temperature $T$
in question, are to be counted. The degeneracy factors $g_f$ and $g_b$ are simply
equal to the number of spin states for each particle (including the spin states
of the antiparticle if it is distinct from the particle). So finally we obtain the expression
for the total pressure exerted on the wall by the plasma inside to be:
\begin{equation}
Pressure =\left( (n_{eff})_f * \frac{7}{8} + (n_{eff})_b \right) *
\frac{2}{3} \sigma T^4 = n_{eff} *\frac{2}{3} \sigma T^4,
\end{equation}
where
\begin{equation}
n_{eff} = \left( (n_{eff})_f * \frac{7}{8} + (n_{eff})_b \right) \label{f149}
\end{equation}
is the total effective number of degrees of freedom.

For example, if we consider a temperature of $T=2$ MeV, the heavy quarks and leptons
are not relevant particles. Even the lightest family $u$ and $d$ quarks are presumably
effectively confined inside nucleons and are not truly present. Also, apart from the massless
photon, the gauge bosons are no longer appreciably present. Thus only electrons and positrons
contribute to the the effective number of degrees of freedom at $T=2$ MeV. The degeneracy
factor for the electron and positron is $g_e = 2*2 =4$ and so we obtain the effective
number of degrees of freedom at $T= 2$ MeV to be:
\begin{equation}
 (n_{eff})_ f \approx \frac{4``\Delta m''}{2\ \hbox{MeV}} * 0.159
\approx \frac{4 * \sqrt{3/4} * m_e}{2 \ \hbox{MeV}} * 0.159 \approx 0.14.
\end{equation}
Here $m_e$ is of course the electron mass.

Finally we can write down an improved formula for the critical size
$R_{crit}(T)$ conceived of as a ball radius,
\begin{equation}
  2S/ R_{crit}(T) = Pressure = n_{eff} *\frac{2}{3} \sigma T^4.
\end{equation}
Thus we get
\begin{equation}
 R_{crit}(T) = \frac{3S}{n_{eff} \sigma T^4}\label{f152}.
\end{equation}

\subsection{Nuclear physics inside the ball}
\label{nuclearphysics}
As the ball contracts after the plasma has gotten so thin that it
cannot carry the pressure of the wall any longer the nucleons
will, if the temperature is under about 10 MeV, be kept inside the
ball and their density will increase significantly.  That will of
course change the balance between bound and free nucleons. In thermodynamic
equilibrium the temperature $T_{NUC}$ at which some bound state having nucleon
number $A$ - say $^4$He - with binding energy $B_A$
occurs in similar amounts as the
nucleons is given by \cite{Kolb}
\begin{equation}
 T_{NUC} = \frac{B_A/(A-1)}{\ln{\eta^{-1}} + 1.5 \ln(m_N/T_{NUC})}\label{NUC}.
\end{equation}
Here $\eta$ is defined as the number of nucleons divided by the
number of photons at the relevant temperature, i.e $\eta =
n/n_{\gamma} \approx n/T^3$.

Before the contraction of the balls, when the total volume inside
the balls was of the same order as the outside ball volume,
the value of the nucleon photon ratio $\eta$ should have been
of the same order as obtained in standard Big Bang Nucleosynthesis
fits. However
{\em those baryons which at the end make up our dark matter balls
were also present at this early stage}. So
rather than the usual fitted value from Big Bang Nucleosynthesis
$\eta \approx 6 * 10^{-10}$, there would have been
about a factor of 7 bigger value $\eta \approx 4
* 10^{-9}$. But now when the balls contracted and the nucleons were
kept inside, while the photons could rather escape, the value of
$\eta$ inside the ball increased drastically. If the temperature
inside the ball got in balance with the outside
even after the contraction of the
ball radius, say from 6 m to 0.67 cm, the $\eta$-parameter would
go up by a factor of $7 * 10^8$.
It would then end up with the value
$\eta \approx 3$. This is what would
happen, if the collapse happened just at the 10 MeV era.
However if for example the contraction first happened,
as we rather expect from (\ref{contraction}), at 2.3 MeV there would be a stronger
contraction by a factor of 4.3
in distance and the final value of $\eta$ would be a factor $4.3^3$ higher
say $\eta \approx 240$. But before reaching
the full contraction of today's dark matter, we expect that nuclear reactions could set in
and make first mainly helium nuclei which then a little later fuse to heavier elements.

For instance at a temperature scale of 2.3 MeV, which is our above estimate
for the typical temperature at contraction of the balls, we achieve the critical
relative density $\eta$ of nucleons to photons from equation (\ref{NUC}) with
$T_{NUC} $ put equal to 2.3 MeV. This gives us the
following equation for the value of $\eta$ at
our supposed contraction era
\begin{equation}
 2.3 \ \hbox{MeV}  = \frac{(28\ \hbox{MeV})/(4-1)}{-\ln{\eta}
 + 1.5 \ln(m_N /2.3\ \hbox{MeV})}.
\end{equation}
Here we have used\footnote{For simplicity we shall use the nuclear binding
energies in the normal vacuum without a bound state condensate.}
the helium binding energy $B_4 =28 \ \hbox{MeV}$. Hence
helium gets thermodynamically favoured at a temperature of 2.3 MeV for
the critical value of $\eta$ given by
\begin{equation}
 \ln{\eta} = 1.5 *6.0  - \frac{28\ \hbox{MeV} }{ 2.3 *3\ \hbox{MeV}}
= 5.0.
\end{equation}
So the helium formation gets thermodynamically favoured once the
contraction has gone so far that $\eta $  becomes about
$\exp(5.0)=150$. This value is indeed reached during the contraction
in as far as the contraction ends up having increased $\eta$
to $240$. A priori the thermodynamical favouring of the helium
does not necessarily mean that the helium is formed immediately.
However we are here talking about a rather high temperature situation
and it may not take long for the helium to form. (This question of the
rate of the transition to the various elements such as helium deserves
further study.)

Helium is just an example of a nucleus that will be formed,
although it can be seen to be the first one getting thermodynamically
favoured as $\eta$ increases. In fact we find the crucial number
$B_A/(A-1)$ in the numerator of formula (\ref{NUC}) to be $28/3
= 9.4$ MeV for helium, while for the competing candidate carbon one
gets $ 92.2/11 = 8.4$ MeV. So, although it is not a convincingly
big difference, the suggestion is that helium formed first and then
the heavier nuclei like carbon later.
Outside a ball, in the ordinary big
bang nucleosynthesis story, you hardly get any carbon at all; carbon is
rather formed much later in the stars.
For example, we find that the
critical value of $\eta$ needed to thermodynamically favour the formation
of carbon becomes
$ \ln{\eta} = 5.4$, which is
to be compared with the value of $ \ln{\eta} = 5.0$ for helium.
This means that the
critical $\eta$ is $\exp(0.4)\approx 1.5$ times bigger for carbon
than for helium.

Although the limits for thermodynamical favouring of the formation
of different size nuclei is not so great measured in terms of the
critical $\eta$, the moments during the contraction when the
formations truly take place could be more different. The point is
that the Coulomb barrier for the formation of heavier nuclei is
larger than for helium. Indeed the height of the Coulomb
barrier is in the range of a few MeV, so that it can be quite
significant at our estimated contraction temperature of 2.3 MeV.

So it is highly suggested that helium forms first and then later
on the heavier elements.   If indeed there is some bottle-neck,
due e.g. to the Coulomb barrier, appreciably delaying the formation
of the heavier elements compared to the helium, then when
finally the heavy elements get formed the process will happen
explosively.
Also by that time the contraction of the balls
may have almost come to the end and thus they may
have the sizes closer to those of
today. At such a late time, with all the balls being already
small, there will no longer be big balls around to take up emitted
nucleons. Under the explosive fusion of helium to heavier nuclei,
the excess energy from the higher binding energy in the heavier
nuclei compared to that in helium is likely to be mainly
transferred to freely moving nucleons, which are more loosely bound into
the ball than the nuclei. These nucleons may then escape from the ball due
to their explosively increased temperature. Once the nucleons are
outside and all the balls have contracted to be small in size, they
will no longer find any ball that can capture them. Thus these
expelled nucleons will stay
around forever in the outside of the balls (i.e. in the phase without
condensate) and actually become the ordinary matter.

In this argument we talked as if neutrons and protons felt completely
the same forces and we just said that nucleons were expelled. One
may, however, become worried that, especially if the ball is already
almost contracted to today's size, there is a Fermi sea of degenerate electrons
extending a bit outside the skin. Thereby a region of electric field is arranged
outside the skin in which the protons are pulled outward due to the excess of protons inside the skin
region. One might then think that only the protons but not the neutrons would mainly be
expelled. This is, however, not true because
 there will develop in the interior of the ball, at whatever stage in size
the explosion occurs, a situation corresponding to a chemical potential for which
the relative density of neutrons and protons in the regions outside the ball
is about unity. This will mean that the nuclei inside the ball
are more neutron rich statistically than if they were in free space.
This effect of a chemical potential, which if you did not count the electric potential
would be much higher for neutrons than for protons inside the ball, is stronger the more contracted
the ball is at the moment we consider. Really we may just say that, due to the positive
electric potential inside the ball and the Boltzmann distribution, there will be fewer protons than neutrons
compared to what would happen without this potential. It is of course also anyway expected that,
if the degenerate electrons became very copious, the whole inside material would have to be
almost pure neutrons like in a neutron star. So the inside matter will be more neutron rich
than in free space. But this fact then means that, also following still the effect of the electric
potential, the nucleons released during the explosive fusion of helium and
running among the nuclei are more often neutrons than protons.
Actually this excess of neutrons over protons corresponds to a chemical potential difference
just equivalent to the electric potential for the protons inside the ball.
Consequently when the nucleons run out of the ball they will reach a neutron proton
ratio close to unity.
So the nucleons coming out of the balls will not deviate much from the equilibrium
distribution usually assumed in estimating the next step of
standard Big Bang Nucleosynthesis.

Actually at an earlier stage of the development of our
present dark matter model \cite{crypto}, we  based a calculation of
the dark matter to ordinary matter ratio on the hypothesis that all
ordinary matter was expelled from dark matter balls in an explosive
fusion of helium nuclei into heavier nuclei. We simply used the
difference in binding energy per nucleon in helium, which is $7.1 \
\hbox{MeV}$, with that in the heavier elements, which is rather $8.5 \
\hbox{MeV}$. Thus during the fusion process there is an excess energy
released per nucleon of $8.5\ \hbox{MeV}  - 7.1 \ \hbox{MeV}  =
1.4\  \hbox{MeV}$, which is just sufficient to release from the
8.5  MeV binding one sixth of the nucleons. This then means that
there is sufficient energy to release approximately one sixth of the
original nucleons. It then follows that the ordinary matter would
make up one sixth of the total amount of matter (both dark and
ordinary matter together). Correspondingly the amount of ordinary matter
would be one fifth of the amount of dark matter.  It should be
remarked that we here ignored the energy needed for the nucleon
to escape though the ball-skin into the outside.
However, since we assume that the typical balls are on the
borderline of expelling their nucleons and collapsing totally
away, the total escape potential - composed from the electric potential,
the potential for crossing the wall and the chemical potential -  will
add up to zero.

This would be exactly true for a ball just on the stability borderline.
However, according to appendix B.2,
the median sized ball has after the
collapse a radius $R_{median}^{(after)}$ that is
larger than the radius $R_{border}^{(after)}$ of the borderline ball by a factor of $2^{4/9}$.
Now the degenerate electron pressure inside the ball is related to
the Fermi energy of the electrons by
(\ref{P}), meaning the pressure
is proportional to the {\em fourth} power
of the electron Fermi-momentum $p_f$.
This pressure is inversely proportional
to the radius of the ball and so
the Fermi momentum of the median sized ball
is $2^{-1/9} $ times the Fermi momentum for the borderline ball.
Now the effective binding energy of the nucleons into
a borderline ball is essentially
zero. Hence the nucleons in a median
sized ball will have an effective binding energy of
$(1-2^{-1/9})\Delta V \approx0.74 \
\hbox{MeV}$.
So the energy needed  per nucleon to
escape from the median sized ball
to the outer space is increased from
7.1 MeV to
(7.1 + 0.74) MeV =$ 7.8_4$ MeV. Thus with
this correction we
rather predict that the amount of
ordinary matter relative to all
the matter (ordinary plus dark) is
1.4 MeV/ (8.5 MeV +0.74 MeV) =1/6.6.
Similarly the amount of ordinary matter
counted relative to the dark matter
is now predicted to be 1.4/(7.1 +0.74)
= 1/5.6.
This to be compared with the recent
Planck satellite result \cite{planck} of
4.9\%/26.4\% = 1/5.44. Accidentally the agreement is better
than the uncertainty in our calculation.

\end{document}